\documentclass{aa}
\usepackage{graphicx}
\usepackage[varg]{txfonts}
\usepackage{longtable}
\usepackage{lscape}
\usepackage{natbib}
\usepackage{url}
\usepackage{color}
\usepackage{multirow,bigdelim}
\usepackage{cases}
\usepackage{amssymb}
\usepackage[colorinlistoftodos]{todonotes}

\bibpunct{(}{)}{;}{a}{}{,} 
\newcommand{\logg}{\log\,g}
\newcommand{\vmic}{\xi_{\rm t}}
\newcommand{\vmac}{\zeta_{\rm mac}}
\newcommand{\vsini}{v\cdot\sin\,i}

\newcommand\kms{{\rm\,km\,s^{-1}}}

\newcommand\teff{T_{\rm eff}}

\newcommand{\uprefs}[1]{$^{#1}$}

\begin{document}

\title{
	Radial abundance gradients in the outer Galactic disk\\ as traced by main-sequence OB stars\thanks{Based on data obtained with the Magellan Clay telescope at the Las Campanas observatory and the ESO/MPI telescope at La Silla under the ESO--ON agreement.}
}
\titlerunning{Radial abundance gradients in the outer Galactic disk}

\author{
G.A.~Bragan\c ca\inst{1}
\and
S.~Daflon\inst{1}
\and
T.~Lanz\inst{2}
\and
K.~Cunha\inst{1,3}
\and
T.~Bensby\inst{4}
\and
P.J.~McMillan\inst{4}
\and
C. D.~Garmany\inst{5}
\and
J.W.~Glaspey\inst{5}
\and
M.~Borges~Fernandes\inst{1}
\and
M.S.~Oey\inst{6}
\and
I.~Hubeny\inst{3}
}
\institute{
Observat\'orio Nacional-MCTIC, Rua Jos\'e Cristino, 77. CEP: 20921-400, Rio de Janeiro, RJ, Brazil\\
\email{daflon@on.br}
\and
 Observatoire de la Cote d'Azur, F-06304 Nice, France
\and
Steward Observatory, University of Arizona, 933 N. Cherry Ave., Tucson, AZ 85721, USA
\and
Lund Observatory, Department of Astronomy and Theoretical physics, Box 43, SE-221\,00 Lund, Sweden
\and
National Optical Astronomy Observatory, 950 N. Cherry Ave., Tucson, AZ 85719, USA
\and
University of Michigan, Department of Astronomy, 311 West Hall, 1085 S. University Ave., Ann Arbor, MI 48109-1107, USA
}


\date{Received XX Xxxxxxx 2018 / Accepted XX Xxxxxxx 2018}

 \abstract{
Elemental abundance gradients in galactic disks are important constraints for models of how spiral galaxies form and evolve. However, the abundance structure of the outer disk region of the Milky Way is poorly known, which hampers our understanding of the spiral galaxy that is closest to us and that can be studied in greatest detail. Young OB stars are good tracers of the present-day chemical abundance distribution of a stellar population and because of their high luminosities they can easily be observed at large distances, making them suitable to explore and map the abundance structure and gradients in the outer regions of the Galactic disk. 
}
 {
Using a sample of 31 main-sequence OB stars located between galactocentric distances $ 8.4 - 15.6$\,kpc, we aim to probe the present-day radial abundance gradients of the Galactic disk.
 }
 {
The analysis is based on high-resolution spectra obtained with the MIKE spectrograph on the Magellan Clay 6.5-m telescope on Las Campanas.  We used a non-NLTE analysis in a self-consistent semi-automatic routine based on \texttt{TLUSTY} and \texttt{SYNSPEC} to determine atmospheric parameters and chemical abundances.
}
 {
Stellar parameters (effective temperature, surface gravity, projected rotational velocity, microturbulence, and macroturbulence) and silicon and oxygen abundances are presented for 28 stars located beyond 9\,kpc from the Galactic centre plus three stars in the solar neighborhood. The stars of our sample are  mostly  on the main-sequence,  with effective temperatures between $20\,800 - 31\,300$\,K, and surface gravities between $3.23 - 4.45$\,dex.
The radial oxygen and silicon abundance gradients are negative and have slopes of $-0.07$\,dex/kpc and $-0.09$\,dex/kpc, respectively, in the region $8.4 \leq R_G \leq 15.6$\,kpc.
 }
 {
The obtained gradients are compatible with the present-day oxygen and silicon abundances measured in the solar neighborhood and are consistent with radial metallicity gradients predicted by chemodynamical models of Galaxy Evolution for a subsample of young stars located close to the Galactic plane.
}
   \maketitle

\section{Introduction}

To understand how large spiral galaxies like the Milky Way formed and evolved to their current state is a major goal in Galactic research. Observational results from analyses of the Galactic structure and chemical composition combined with modern models of Galactic chemical evolution are required to
constrain how galaxies form and evolve  \citep{maiolino2018}. However, the observational picture of
the Milky Way is incomplete and still lacks some major pieces. In particular the outer disk is poorly mapped and poorly understood. This is unfortunate, as Galactic radial gradients of elemental abundances are important observational
constraints for models of Galactic chemical evolution  \citep{maiolino2018, wang2019}. These gradients are influenced by the evolution of the Milky Way on the largest scales. Differences in the history of the halo evolution may affect the chemical distribution in the outer disk at galactocentric distances greater than 10\,kpc, whereas the inner gradients
remain unchanged \citep[e.g.,][]{chiappini2001}.

Massive stars are considered the main drivers of the Galactic chemo-dynamical evolution  \citep{crowther2012, matteucci2008}. They are the first stars to process hydrogen to heavier elements and to return these processed elements to the interstellar medium.  Their strong winds and supernova explosions are important triggers of star formation  \citep{crowther2012}. Also, due to their short lifetimes, OB stars are naturally bound to their place of birth in the Galactic plane \citep{schulz2012}. Thus they provide observational constraints on the present state of the Galaxy through detailed analyses of their photospheres. The abundance radial gradients derived from these objects can be used to infer the infall rate and mixing of the gas within the thin
disk  \citep{chiappini2001}.

Massive stars have been studied for decades by several groups using photometric and spectroscopic tools \citep[e.g.,][]{rolleston2000,daflon2004}. However, the samples studied in previous works concentrate in the ``local''
disk and typically include less than about ten stars outside the solar circle.

In this work we present stellar parameters (effective
temperature, surface gravity and $\vsini$, microturbulent and macroturbulent velocities) and oxygen and silicon abundances in non-LTE for a sample of 28 main-sequence OB stars located towards the Galactic anti-center, between galactocentric radii $\sim$ 9.5 and 15.6\,kpc, plus three stars in the solar neighbourhood. The sample analyzed here is part of the larger sample of 136 OB stars in the outer disk analysed by \cite{garmany2015}, where multiplicity, photometric effective temperature, and projected rotational velocity ($\vsini$) were determined.

This paper is structured as follows: Section~\ref{sec:observation} describes the target selection and the observations; Sect.~\ref{sec:analysis} explains the methodology used to analyse the stellar spectra and Sect.~\ref{sec:distances} explains how the distances of the stars were obtained. In Sect.~\ref{sec:results} we discuss the determination of stellar parameters and abundances, and in Sect.~\ref{sec:gradients} we discuss the radial distribution of the silicon and oxygen abundances; our conclusions are presented in Sect.~\ref{sec:conclusion}.

\section{Observations and sample selection}
\label{sec:observation}

\begin{figure}
\resizebox{\hsize}{!}{
\includegraphics{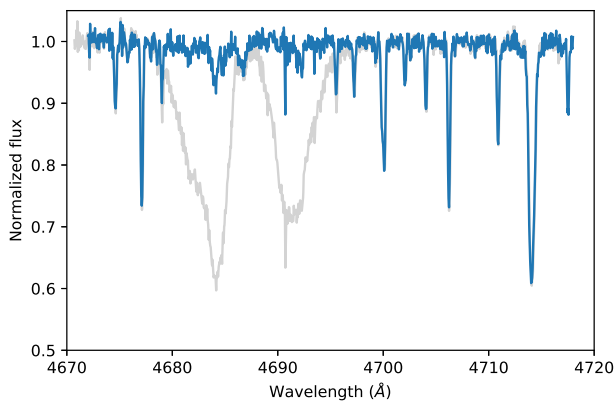}}
\caption{
 Example spectra in the region $4670-4720$\,{\AA} for the star ALS 14013 reduced with the pipeline (black line) and re-reduced (red line). The feature introduced by the CCD defect in the region $4680-4695$\,{\AA} has been adequately removed. 
\label{fig:pattern}
}
\end{figure}

\begin{figure}
\resizebox{\hsize}{!}{
\includegraphics{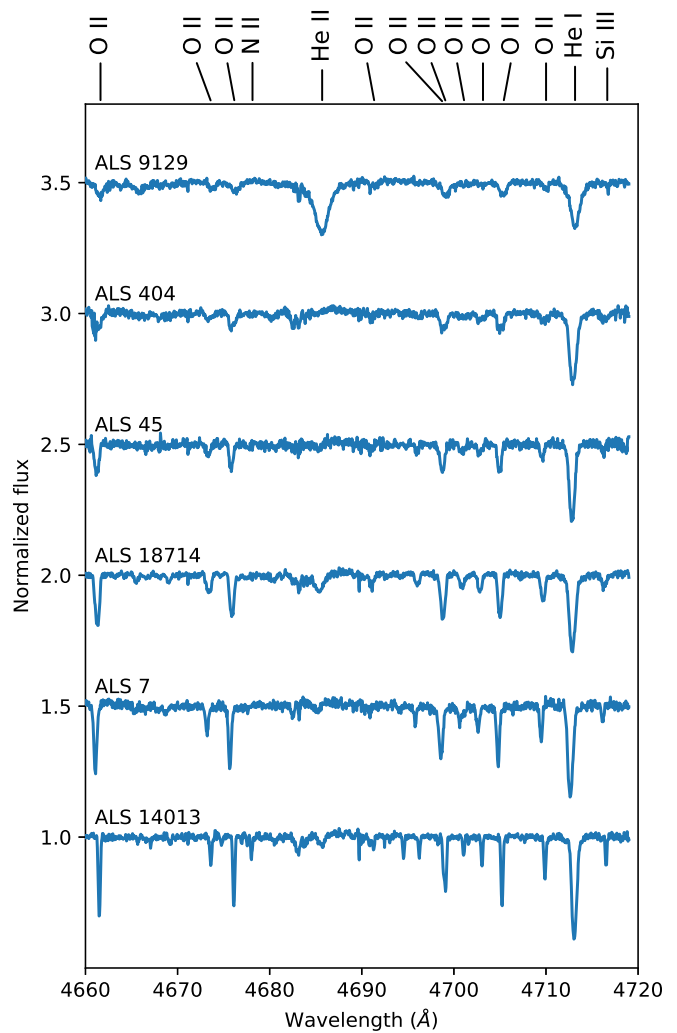}}
\caption{Example spectra for  six observed stars in the region $4660-4720$\,{\AA}. The spectra are  vertically shifted for better visualization. Some spectral lines are identified on the top. \label{fig:spectra}
}
\end{figure}

\begin{figure}
\resizebox{\hsize}{!}{
\includegraphics{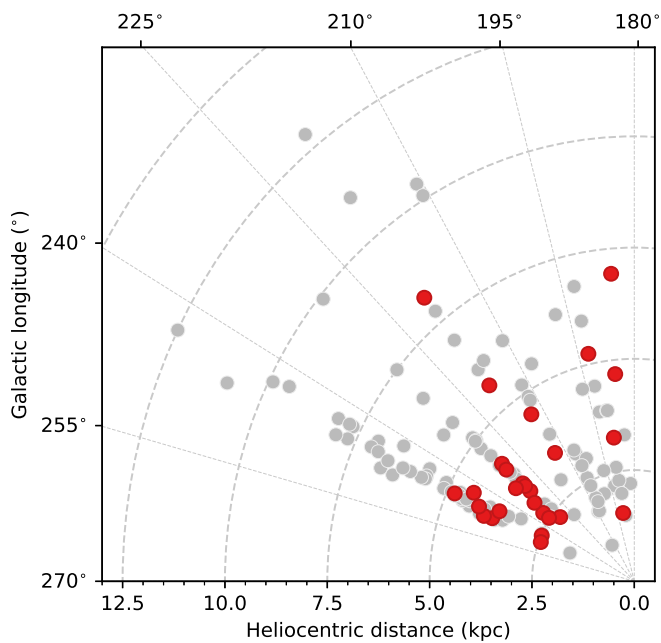}}
\caption{Positions of the all 136 observed stars (grey circles) projected onto	the Galactic plane; the 28 stars selected for a detailed abundance	analysis are presented (red circles) with reviewed distances (see Sect.~\ref{sec:distances}). All stars are located in the third Galactic quadrant. \label{fig:polar}}
\end{figure}

The target stars were selected from the catalogues by \cite{reed1998,reed2003}, using the reddening-free index Q  to select stars with spectral types earlier than B2.
Additional selection criteria include the stellar magnitudes ($V>10$) and the location towards the outer edge of the Galactic disk in order to provide OB stars that are likely to be located at large galactocentric distances.
A detailed description of the selection criteria of the sample stars is presented in \cite{garmany2015}.

High-resolution spectra were obtained over six nights (Dec 29, 2007 to Jan 3, 2008) for 136 stars with the MIKE spectrograph \citep{bernstein2003} at the Magellan Clay 6.5\,m telescope at Las Campanas Observatory in Chile. MIKE is a double echelle spectrograph that simultaneously records red ($4900-9500$\,{\AA}) and blue ($3350-5000$\,{\AA}) spectra.  The observations were done with a 0.7\,arcsec slit, providing  spectra with nominal resolution of about $R\approx 53,000$ with the blue arm, as measured in the Th-Ar calibration images.
The exposure times were estimated to give spectra with S/N $\sim$ 100 per pixel for the stars that have V magnitudes in the range $10 - 14$\,mag.

At the time of the observing run, the blue CCD of the MIKE spectrograph was suffering from a serious disfiguring feature affecting a number of orders, and the CCD response function was changing with time (Ian Thompson, private communication; the CCD was replaced in 2009). In addition, a strong contamination in the line profiles was introduced  by the use of a B
star as a milky flat. As a result, the spectra that were reduced using the standard pipeline displayed a relatively strong feature (about $20-30$\,\%) in the central regions ($\sim$20\AA) of the orders 1-12 of the blue arm (i.e.,
$5000-4400$\,{\AA}). Thus, the spectra had to be re-reduced using the data reduction packages written in Python and provided by Carnegie Observatories\footnote{The Carnegie Observatories CarPy python distribution is available for download at \url{http://code.obs.carnegiescience.edu/carnegie-python distribution}.}. This resulted in better quality spectra (corrected for the blaze function and CCD defect) obtained through a careful flat-field selection, using quartz lamp spectra, as shown in Fig.~\ref{fig:pattern} for  the star ALS 14013, in the region $4680-4695$\,{\AA}. 

Some intrinsic characteristics of early-type stars such as fast rotation, binarity, and stellar winds, can contribute to make the spectral analysis of massive OB stars challenging.
Stellar rotation ($\vsini$) becomes a dominant broadening mechanism and affects in particular the weak metal lines present in the spectra of OB stars; these are smeared at $\vsini \approx 100 \kms$ or higher, preventing an accurate abundance analysis. In this study, we selected only those stars with sharp-lined spectra (with $\vsini < 70 \kms$) in order to avoid any systematics and larger errors that could be introduced in our abundance results due to high stellar rotation. Recently, \cite{cazorla2017} analyzed two fast-rotating stars (ALS 864 and ALS 18675) from our full sample. 

A majority of massive stars are found in binary or multiple star systems \citep{sana2012,sana2013} and some of them may present double (or even multiple) lines in their spectra (double-lined spectroscopic binaries, SB2).
In such cases one has to disentangle the component spectra
\cite[e.g.][]{harmanec2004} and treat the resulting  spectra
individually. Massive stars with moderate to strong winds may also display emission lines in the center of the hydrogen lines. Depending on the intensity of the emission lines, the wings of hydrogen lines that are generally used to determine the surface gravity may be affected.

From the original sample of 136 stars described in \cite{garmany2015}, stars with broad line profiles ($\vsini > 70  \kms$), strong emission profiles in the center of H lines,  and clear double lines were eliminated for this study. One additional star (ALS 16807) was also discarded  (see Section 3.2).

Single-epoch spectra like those analyzed in this paper are not adequate to identify and study binary systems. However, in order to discard potential single-lined binaries, we did a search in the literature for radial velocity measurements or any information on binarity. The results are the following:
\begin{itemize}
    \item 	For the stars ALS 45, ALS 404, ALS 428, ALS 9209, ALS 18679, and ALS 18681, we compared the observed spectra with previous data obtained with FEROS @ ESO 1,52m by our group. Only ALS 428 shows significant variation of radial velocity, although no spectral line from a companion could be detected in the two spectra. 
    
    \item 	For the stars ALS 384, ALS 9209, and ALS 18714 we compared our measurements of radial velocity with previous values listed in the literature: the two sets of measurements are consistent within 10 km/s. 
    
    \item 	The star ALS 17694 has been studied by \cite{lester2016} as a possible member of an eclipsing binary and has been discarded as a component of the system.
    
    \item For the remaining 19 stars, no measurement of radial velocity and no information on binarity are available in the literature. We thus adopted a simple approach based on visual inspection of single-epoch spectra searching for potential asymmetry in the line profiles or any other spectral signature from a companion. As no clear signature has been found in the observed spectra, we considered these stars as single and we proceeded the abundance analysis. 
\end{itemize}

Our final sample contains 28 stars suitable for a detailed chemical abundance analysis. Figure~\ref{fig:spectra} shows sample spectra in the  region between 4660 \AA \ and 4720 \AA \  for six stars in our sample. The positions of all observed stars projected in the Galactic plane are shown in Fig.~\ref{fig:polar}. Grey circles show the original sample of 136 stars and red circles show the subsample of 28 stars
selected for the abundance analysis with revised heliocentric distances as described in Sec. \ref{sec:distances}.

\section{Abundance analysis}
\label{sec:analysis}

\subsection{Non-LTE analysis and atomic data}

Atmospheric parameters and  chemical  abundances were determined using a grid of non-LTE line-blanketed model atmospheres calculated with \texttt{TLUSTY}
\citep{hubeny1988,hubeny1995, hubenylanz2017} and the synthesis code \texttt{SYNSPEC}.
The model atoms used to compute the original atmosphere grids OSTAR2002 \citep{lanz2003} and  BSTAR2006 \citep{lanz2007} have been updated: in the place of superlevels previously adopted, the updated model atoms now  explicitly include the higher energy levels. The updated models of \ion{C}{ii}, \ion{C}{iii}, and \ion{C}{iv} have 60, 95, and  55 energy levels, respectively. The models of \ion{O}{i} and \ion{O}{ii} have been updated to include  69  and  219 levels, respectively, while the updated models of \ion{Si}{ii}, \ion{Si}{iii}, and \ion{Si}{iv} have 70, 122, and  53 levels, respectively.
A new grid of model atmospheres was created with 20 values of
effective temperatures in the range between $14\,000-33\,000$\,K, in steps of 1000\,K, and 13 values of surface gravity in the range between $3.0-4.5$\,dex, in steps of 0.12\,dex. All models have a microturbulent velocity of $2.0\,\kms$.

Stellar parameters and abundances of oxygen and silicon were determined using the list of lines ( \ion{O}{i}, \ion{O}{ii}, \ion{Si}{ii},  \ion{Si}{iii}, and \ion{Si}{iv}) presented in Table~\ref{tab:atomic_data} with their respective $\log gf$-values.

\begin{table}
\centering
\caption{Line list and adopted values of oscillator strength ($\log gf$).
	\label{tab:atomic_data}
}
\footnotesize
\setlength{\tabcolsep}{5mm}
\begin{tabular}{clcc}
\hline\hline
\noalign{\smallskip}
Wavelength       &
Specie           &
$\log gf$        &
Reference        \\
(\AA)            &
                 &
                 &
		 \\
\noalign{\smallskip}
\hline
\noalign{\smallskip}
  7771.94 & O I    & \phantom{-}0.354 & 3 \\
\noalign{\smallskip}
\hline
\noalign{\smallskip}
  4062.91 & O II   & -0.090           & 1 \\
  4071.24 & O II   & -0.090           & 1 \\
  4071.24 & O II   & -0.090           & 1 \\
  4078.84 & O II   & -0.287           & 3 \\
  4083.93 & O II   & \phantom{-}0.150 & 1 \\
  4085.11 & O II   & -0.191           & 3 \\
  4089.29 & O II   & \phantom{-}0.892 & 1 \\
  4092.93 & O II   & -0.325           & 3 \\
  4119.22 & O II   & \phantom{-}0.447 & 3 \\
  4129.32 & O II   & -0.945           & 3 \\
  4132.80 & O II   & -0.067           & 3 \\
  4590.97 & O II   & \phantom{-}0.331 & 3 \\
  4595.96 & O II   & -1.033           & 1 \\
  4596.18 & O II   & \phantom{-}0.180 & 3 \\
  4609.37 & O II   & \phantom{-}0.670 & 1 \\
  4610.17 & O II   & -0.170           & 1 \\
  4638.86 & O II   & -0.325           & 3 \\
  4641.81 & O II   & \phantom{-}0.066 & 3 \\
  4649.14 & O II   & \phantom{-}0.324 & 3 \\
  4661.63 & O II   & -0.268           & 3 \\
  4669.27 & O II   & -1.030           & 1 \\
  4669.47 & O II   & -0.610           & 1 \\
  4701.18 & O II   & \phantom{-}0.088 & 1 \\
  4701.71 & O II   & -0.611           & 1 \\
  4703.16 & O II   & \phantom{-}0.262 & 2 \\
  4705.35 & O II   & \phantom{-}0.533 & 3 \\
  4710.01 & O II   & -0.090           & 3 \\
  4906.83 & O II   & -0.157           & 3 \\
  4941.07 & O II   & -0.018           & 3 \\
  4943.00 & O II   & \phantom{-}0.307 & 3 \\
\noalign{\smallskip}
\hline
\noalign{\smallskip}
  5591.66 & O III  & -1.990           & 3 \\
  5592.25 & O III  & -0.361           & 1 \\
\noalign{\smallskip}
\hline
\noalign{\smallskip}
  4130.87 & Si II  & -0.824           & 1 \\
  4130.89 & Si II  & \phantom{-}0.476 & 1 \\
\noalign{\smallskip}
\hline
\noalign{\smallskip}
  4552.62 & Si III & \phantom{-}0.283 & 1 \\
  4554.00 & Si III & -0.202           & 1 \\
  4567.84 & Si III & \phantom{-}0.060 & 1 \\
  4574.76 & Si III & -0.418           & 1 \\
  4813.33 & Si III & \phantom{-}0.706 & 1 \\
  4819.71 & Si III & \phantom{-}0.823 & 1 \\
  4819.81 & Si III & -0.353           & 1 \\
  4828.95 & Si III & \phantom{-}0.938 & 1 \\
  4829.11 & Si III & -0.354           & 1 \\
  4829.21 & Si III & -2.153           & 1 \\
\noalign{\smallskip}
\hline
\noalign{\smallskip}
  4088.86 & Si IV  & \phantom{-}0.200 & 1 \\
  4116.10 & Si IV  & -0.110           & 1 \\
\noalign{\smallskip}
\hline
\end{tabular}
\tablebib{
	(1)~\cite{cunto1992}; (2) \cite{wiese1996}; (3)	\cite{froesefischer2006}
}
\end{table}

\subsection{Stellar atmospheric parameters}
\label{sec:parameters}

\begin{figure}
\resizebox{\hsize}{!}{
\includegraphics{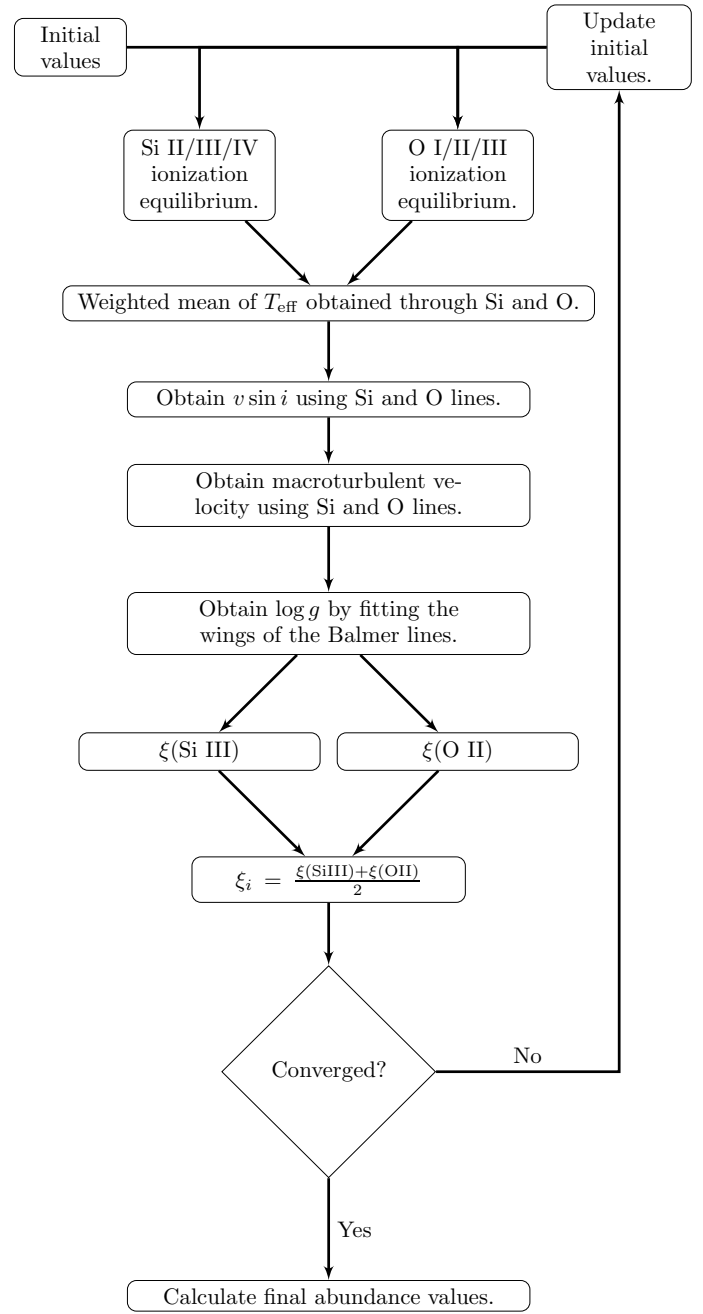}}
\caption{Flowchart illustrating the iterative methodology to determine the stellar parameters and elemental abundances, based on the ionisation	balance of \ion{O}{i}/\ion{O}{ii}/\ion{O}{iii} and	\ion{Si}{ii}/\ion{Si}{iii}/\ion{Si}{iv} and $\chi^2$ minimisation.
\label{fig:flowchart}}
\end{figure}

Spectral features are typically sensitive to more than one atmospheric parameter at the same time: most of the spectral lines show sensitivity to the effective temperature, $\logg$, and microturbulence. For example, the broad wings of hydrogen lines in OB stars are mainly sensitive to the surface gravity but are also (although to a lesser degree) sensitive to the effective temperature. The intermediate and strong metal lines are, as expected, very sensitive to the microturbulent velocity ($\vmic$), whilst the weak lines remain constant with respect to the $\vmic$-value. The projected rotational velocity, $\vsini$, on the other hand, will affect the metal lines in a similar way but the weak lines will not be measurable. All these relations introduce an interdependency in the spectroscopic determination of the stellar parameters.

In our approach, the atmospheric parameters and elemental abundances for the sample stars were determined via an iterative scheme, inspired by the method described in  \cite{hunter2007}. Effective temperatures ($\teff$), surface gravities ($\logg$), projected rotational velocities ($\vsini$), macroturbulent velocities ($\vmac$),  microturbulent velocities ($\vmic$) and abundances of O and Si were determined in a self-consistent way, based on the non-LTE synthesis of line profiles, with the best fits to the observed spectra obtained using $\chi^2$ minimisation. The adopted methodology is illustrated by the flowchart in Fig.~\ref{fig:flowchart}, with the following steps:

\begin{enumerate}

\item Initial estimates of $\teff$ are based on the spectral types; the initial value of $\logg$ is 4.0 dex (a reasonable $\logg$ value for OB stars with luminosity class IV-V).  The  initial  $\vmic$ value is set to $2\,\kms$, the same value as adopted in the calculation of the model atmospheres with \texttt{TLUSTY}.  The main role of $\vmic$, however,  is to promote the broadening of the line profiles but it doe not affect the model convergence \citep{hubenylanz2017b}. For $\vsini$, the values listed in \cite{garmany2015} were adopted as initial guesses.

\item $\teff$ is obtained by requiring ionisation balance of the available species  \ion{Si}{ii}/\ion{Si}{iii}/\ion{Si}{iv} and	\ion{O}{i}/\ion{O}{ii}/\ion{O}{iii}. The new $\teff$ is the average	value weighted according to the number of species available in the	observed stellar spectrum.

\item The fitting of all O and Si lines is repeated for the fixed $\teff$ value	in order to determine  $\vsini$, which is the mean of a gaussian fit to	the distribution of the $\vsini$ values.  Limb darkening coefficient is fixed as $\epsilon=0.6$, following \cite{hubenylanz2017}.

\item A similar fitting procedure is repeated in order to obtain $\vmac$,  adopting a radial-tangential approach \citep{hubenylanz2017}.

\item The surface gravity is determined by fitting the wings of Balmer lines,	which are very sensitive to pressure broadening.  Profiles of	H$\alpha$, H$\beta$ and H$\gamma$ are used whenever available, and the	adopted value is the average from the values obtained from each of the	hydrogen lines.

\item The determination of the microturbulent velocity, $\vmic$, follows the classical approach of requiring the agreement between weak and strong lines. Equivalent widths of \ion{Si}{iii} and \ion{O}{ii} lines are used as indicators of the line strength. The  adopted  stellar microturbulence is the average of $\vmic$-values obtained for  Si and O, rounded to the nearest integer.

\end{enumerate}

Once convergence between input and output values of all the stellar parameters is achieved, the adopted parameters are fixed and  used to refit the spectral lines  to derive the final elemental abundances. We note that for one star (ALS~ 18607) we were not able to obtain convergence using this methodology and this will be further explored in a future study. The stellar parameters and chemical abundances obtained for the 28 stars of our sample are given Table~\ref{tab:sample_results}, in order of increasing ALS number. 
The abundances represent average values computed from the individual lines and the standard deviation.  

\begin{table*}
\centering
\caption{Stellar parameters and elemental abundances for the sample stars.
\label{tab:sample_results}
}
\footnotesize
\setlength{\tabcolsep}{3mm}
\begin{tabular}{@{\extracolsep{4pt}}lcccccccc@{}}
\hline\hline
\noalign{\smallskip}
ID          &
$\teff$     &
$\log g$    &
$v_r$       &
$\vmic$     &
$\vsini$   &
$\zeta$     &
A(O)        &
A(Si) \\
\cline{4-7}
\cline{8-9}
\noalign{\smallskip}
                         &
K                        &
(cgs)                    &
\multicolumn{4}{c}{km/s} &
\multicolumn{2}{c}{dex}  \\
\noalign{\smallskip}
\hline
\noalign{\smallskip}
\multicolumn{9}{l}{Outer disk sample stars} \\
\noalign{\smallskip}
\hline
\noalign{\smallskip}
     ALS 7 &  24,500 & 3.68  & 28 & \phantom{1}7 & 10  & 11  & 8.63 $\pm$0.06 & 7.27 $\pm$0.05 \\
    ALS 45 &  23,700 & 3.83 & 38 & \phantom{1}5 & 18 & 17  & 8.40 $\pm$0.08 & 7.12 $\pm$0.06 \\
   ALS 208 &  26,800 & 4.08 & 70 & \phantom{1}3 & 21 & \phantom{1}7  & 8.70 $\pm$0.05 & 7.30 $\pm$0.07 \\
   ALS 384 &  23,500  & 3.85 & 47 & \phantom{1}3 & 25  & 13   & 8.66 $\pm$0.09 & 7.32 $\pm$0.07 \\
   ALS 404 &  22,500  & 3.77 & 47 & \phantom{1}3 & 38  & 17 & 8.65 $\pm$0.10 & 7.44 $\pm$0.09 \\
   ALS 428 &  25,800  & 4.07 & 57 & \phantom{1}3 & 34  & 46  & 8.51 $\pm$0.05 & 7.22 $\pm$0.07 \\
   ALS 505 &  31,300  & 4.25  & 79 & \phantom{1}6 & \phantom{1}8  & \phantom{1}9 & 8.52 $\pm$0.05 & 7.24 $\pm$0.06 \\
   ALS 506 &  25,700  & 4.03  & 46 & \phantom{1}3 & 18  & 14  & 8.34 $\pm$0.05 & 7.48 $\pm$0.06 \\
   ALS 510 &  24,300  & 3.95 & 66  & \phantom{1}2 & 31  & 12    & 8.54 $\pm$0.07 & 7.41 $\pm$0.06 \\
   ALS 634 &  21,400  & 3.77 & 52  & \phantom{1}5 & 33  & 34    & 8.66 $\pm$0.11 & 7.23 $\pm$0.09 \\
   ALS 644 &  24,800 & 4.15 & 63 & \phantom{1}4 & \phantom{1}6 & \phantom{1}6  & 8.70 $\pm$0.08 & 7.34 $\pm$0.06 \\
   ALS 777 &  21,100  & 3.72 & 54 & \phantom{1}4 & 55   & 11 & 8.51 $\pm$0.12 & 7.23 $\pm$0.08 \\
   ALS 904 &  23,300 & 3.65 & 56 & \phantom{1}5 & \phantom{1}0  & \phantom{1}0 & 8.49 $\pm$0.08 & 7.14 $\pm$0.05 \\
   ALS 914 &  28,100 & 3.92 & 42 & \phantom{1}6 & 20   & 22  & 8.68 $\pm$0.04 & 7.47 $\pm$0.07 \\
   ALS 921 &  24,100 & 3.23 & 56 & 10        & 33  & 48   & 8.75 $\pm$0.08 & 7.45 $\pm$0.12 \\
  ALS 9209 &  28,200  & 4.00 & 62 & \phantom{1}7 & \phantom{1}5 & 12   & 8.58 $\pm$0.03 & 7.19 $\pm$0.04 \\
 ALS 14007 &  22,100  & 3.82 & 74 & \phantom{1}3 & 31 & 25 & 8.70 $\pm$0.10 & 7.32 $\pm$0.07 \\
 ALS 14013 &  24,800  & 3.88 & 56  & \phantom{1}4 & \phantom{1}3  & \phantom{1}2 & 8.59 $\pm$0.07 & 7.32 $\pm$0.05 \\
 ALS 15608 &  26,000 & 3.83 & 70 & \phantom{1}5 & \phantom{1}7  & \phantom{1}6 & 8.58 $\pm$0.03 & 7.18 $\pm$0.04 \\
 ALS 16106 &  24,800  & 4.32 & 74 & \phantom{1}3 & \phantom{1}8  & \phantom{1}8 & 8.59 $\pm$0.08 & 7.22 $\pm$0.07 \\
 ALS 17694 &  30,300  & 4.45 & 30 & \phantom{1}5 & \phantom{1}0 & \phantom{1}0 & 8.56 $\pm$0.03 & 7.15 $\pm$0.03 \\
 ALS 18020 &  29,500  & 4.32 & 63 & 12          & 10  & \phantom{1}0 & 8.62 $\pm$0.03 & 7.26 $\pm$0.02 \\
 ALS 18674 &  29,800 & 4.30 & 68 & \phantom{1}6 & \phantom{1}8 & \phantom{1}9 & 8.30 $\pm$0.04 & 6.92 $\pm$0.03 \\
 ALS 18679 &  30,000  & 4.37  & 66 & \phantom{1}8 & 64  & 22   & 8.25 $\pm$0.04 & 6.85 $\pm$0.05 \\
 ALS 18681 &  28,700  & 4.28 & 82 & \phantom{1}6 & 49   & \phantom{1}1  & 8.32 $\pm$0.04 & 6.88 $\pm$0.05 \\
 ALS 18714 &  27,500 & 4.25 & 43 & \phantom{1}5 & 22   & 10  & 8.69 $\pm$0.04 & 7.38 $\pm$0.07 \\
 ALS 19251 &  24,200  & 3.62  & 38 & \phantom{1}3 & 42  & 11 & 8.49 $\pm$0.06 & 7.14 $\pm$0.09 \\
 ALS 19264 &  20,800  & 3.97 & 32 & \phantom{1}5 & 22  & \phantom{1}5 & 8.51 $\pm$0.11 & 7.13 $\pm$0.07 \\
\noalign{\smallskip}
\hline
\noalign{\smallskip}
\multicolumn{9}{l}{Local disk sample stars} \\
\noalign{\smallskip}
\hline
\noalign{\smallskip}
  HD 61068 &  24,900 & 4.00 & 34 & \phantom{1}5 & 11  & 15  & 8.72 $\pm$0.08 & 7.40 $\pm$0.06 \\
  HD 63922 &  30,700 & 3.82 & 22 & 14   & 27   & 27   & 8.75 $\pm$0.07 & 7.48 $\pm$0.12 \\
  HD 74575 &  21,900 & 3.44 & 14 & \phantom{1}9 & \phantom{1}6 & 20 & 8.77 $\pm$0.10 & 7.45 $\pm$0.08 \\
\noalign{\smallskip}
\hline
\end{tabular}
\end{table*}

\subsection{Uncertainties}
\label{sec:uncertainties}

Uncertainties in the derived abundances were estimated by recomputing the elemental abundances after changing the stellar parameters ($\teff$, log g, $\vsini$, microturbulent and macroturbulent velocities), one at a time, by the uncertainty in the given parameter. The adopted uncertainties in stellar parameters are: $\Delta\teff = 1000$\,K, $\Delta\logg = 0.15$, $\Delta\vsini=15$\,\%, $\Delta\vmic = 2\,\kms$, and $\Delta\vmac=15$\,\%; these represent the precision of the adopted methodology in obtaining each individual parameter.
The total abundance error was computed by adding the individual abundance variations in quadrature. Typical abundance errors range from 0.01\,dex to 0.13\,dex, with a mean of 0.06\,dex.  

The stellar parameters that have the highest impact on the derived O and Si abundances are the microturbulent velocity and the effective temperature. In particular, the \ion{Si}{iii} lines analyzed are good indicators for deriving of the microturbulent velocity since the intermediate-to-strong \ion{Si}{iii} lines are mainly sensitive to the microturbulent velocity: a variation of $2\,\kms$ in $\xi$ results in a difference in average of 0.08\,dex in the Si abundance. The \ion{O}{ii} lines, which are overall weaker than the \ion{Si}{iii} lines, are less sensitive to the microturbulent velocity: for oxygen, the same average difference is 0.05\,dex. The adoption of $\Delta \teff  = 1000$~K results in a average variation of 0.11 and 0.06 for oxygen and silicon respectively, showing that \ion{O}{ii} lines are sensitive monitors of the effective temperature. For most of the stars studied, the total random errors in the elemental abundances are generally below 0.06\,dex.

\subsection{Consistency checks}
\label{sec:check}

We tested our methodology on three well-studied OB stars from the literature which are located closer to the Sun (compared to the outer disk sample): HD\,61068, HD\,63922 and HD\,74575.
We note that HD\,63922 is a known binary, although we do not see any spectral signature from the secondary, as also observed by \cite{nieva2012}. The analysis was based on high-quality spectra  ($S/N \approx 400$) obtained with the FEROS spectrograph \citep{kaufer99} on the ESO/MPI 2.2m telescope. The derived stellar parameters and oxygen and silicon abundances are listed in the bottom part of Table~\ref{tab:sample_results}, and these three stars are referred to as ``Local disk sample stars''.

 In Table~\ref{tab:comp_results} we compare the results obtained for the Local disk sample stars in this work (US) and by \cite{nieva2007,nieva2012} (NP). Generally, we find reasonable agreement with the results presented in this table, although with some  differences: our values of $\teff$, $\logg$, $\vsini$ and $\vmac$ tend to be slightly lower ($<6.5$\,\%, $<$0.22 dex, $2-5\,\kms$ and $0-10\,\kms$, respectively), whilst our microturbulent velocities are usually higher ($2-6\,\kms$).   The line broadening parameters seem to compensate each other so that the differences observed in the obtained parameters translate into  abundance variations that are less than 0.04\,dex for oxygen and less than 0.13\,dex for silicon. The difference in the silicon abundance is probably due to our higher values of microturbulent velocity.
Nonetheless, these abundance differences between the two sets of results can be explained in terms of the uncertainties in the abundance determinations (see Sect.~\ref{sec:uncertainties} below). Our conclusion from this comparison is that our  abundance results are  consistent and in good agreement with those of \cite{nieva2007,nieva2012}.


\begin{table*}
\centering
\caption{Comparison between stellar parameters and abundances from this work  (US) and  \cite{nieva2007,nieva2012} (NP)
\label{tab:comp_results}}
\footnotesize
\setlength{\tabcolsep}{3mm}
\begin{tabular}{@{\extracolsep{4pt}}rcccccccc@{}}
\hline\hline
\noalign{\smallskip}
ID          &
$\teff$     &
$\log g$    &
$\vmic$     &
$\vsini$   &
$\zeta$     &
A(O)        &
A(Si) \\
\cline{4-6}
\cline{7-8}
\noalign{\smallskip}
                         &
K                        &
(cgs)                    &
\multicolumn{3}{c}{km/s} &
\multicolumn{2}{c}{dex}  \\
\noalign{\smallskip}
\hline
\noalign{\smallskip}
 HD 61068 (US) &  24,900 & 4.00  & 5 & 11  & 15 & 8.72  & 7.40 \\
          (NP) &  26,300 & 4.15  & 3 & 14 & 20 & 8.76 & 7.53 \\
        (US-NP)& $-$1,600 & $-$0.15 & 2 & $-$3 & $-$5 & $-$0.04 & $-$0.13 \\          
  \hline
 HD 63922 (US) &  30,700  & 3.82 & 14 & 27 & 27  & 8.75 & 7.48  \\
          (NP) &  31,500 & 3.95 & 8 & 29 & 37  & 8.79 & 7.49  \\
       (US-NP) & $-$800 & $-$0.13 & 6 & $-$2 & $-$10 & $-$0.04 & $-$0.01 \\
  \hline
 HD 74575 (US) &  21,900  & 3.44  & 9 & 6  & 20  & 8.77  & 7.45  \\
          (NP) &  22,900 & 3.66  & 5 & 11  & 20  & 8.79  & 7.52  \\
       (US-NP) & $-$1,000 & $-$0.22 & 4 & $-$5 & 0 & $-$0.02 & $-$0.07 \\
\noalign{\smallskip}
\hline
\end{tabular}
\end{table*}

\subsection{Comparisons with other studies}

\begin{figure*}
\centering
\resizebox{\hsize}{!}{
\includegraphics{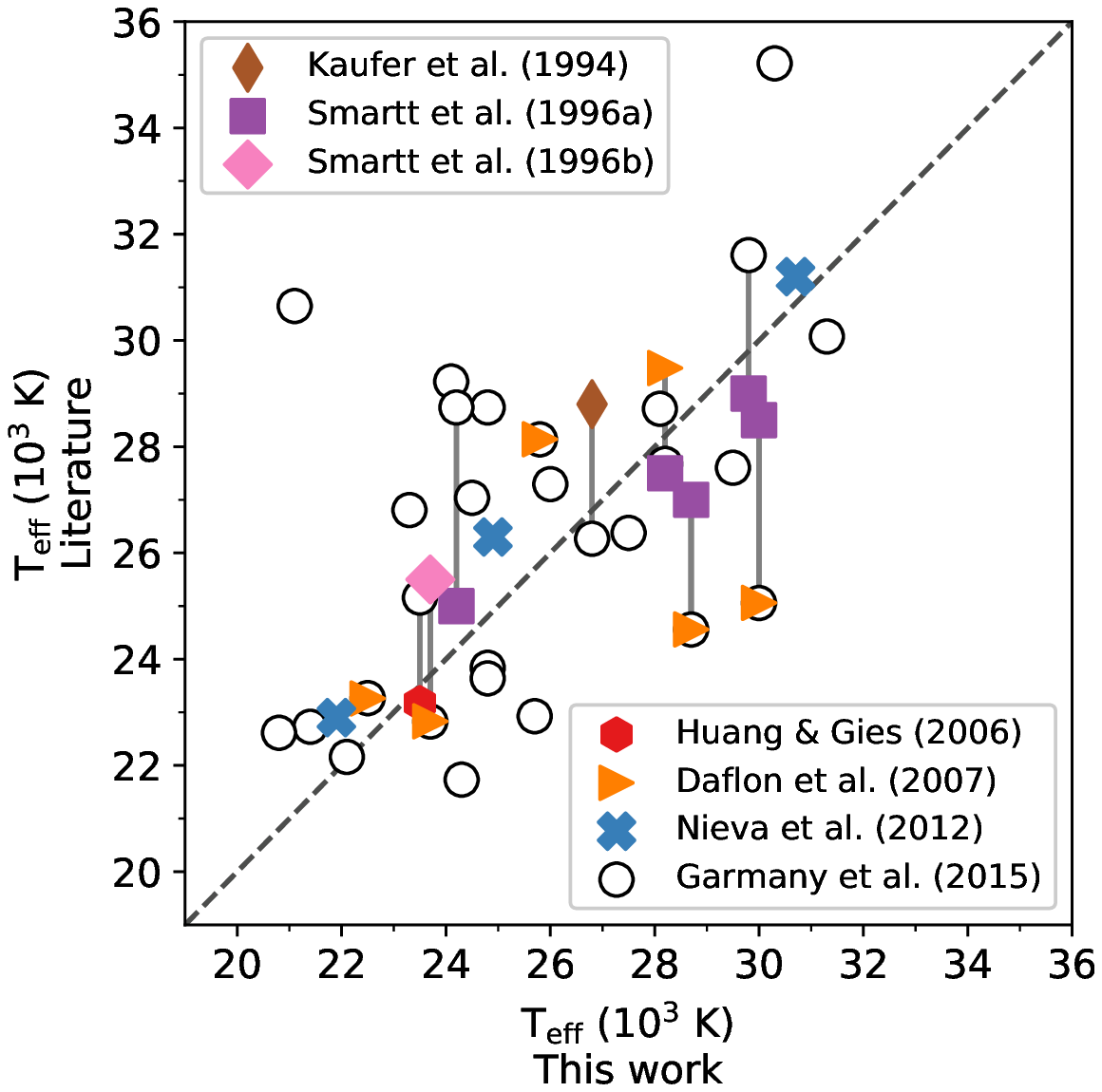}
\includegraphics{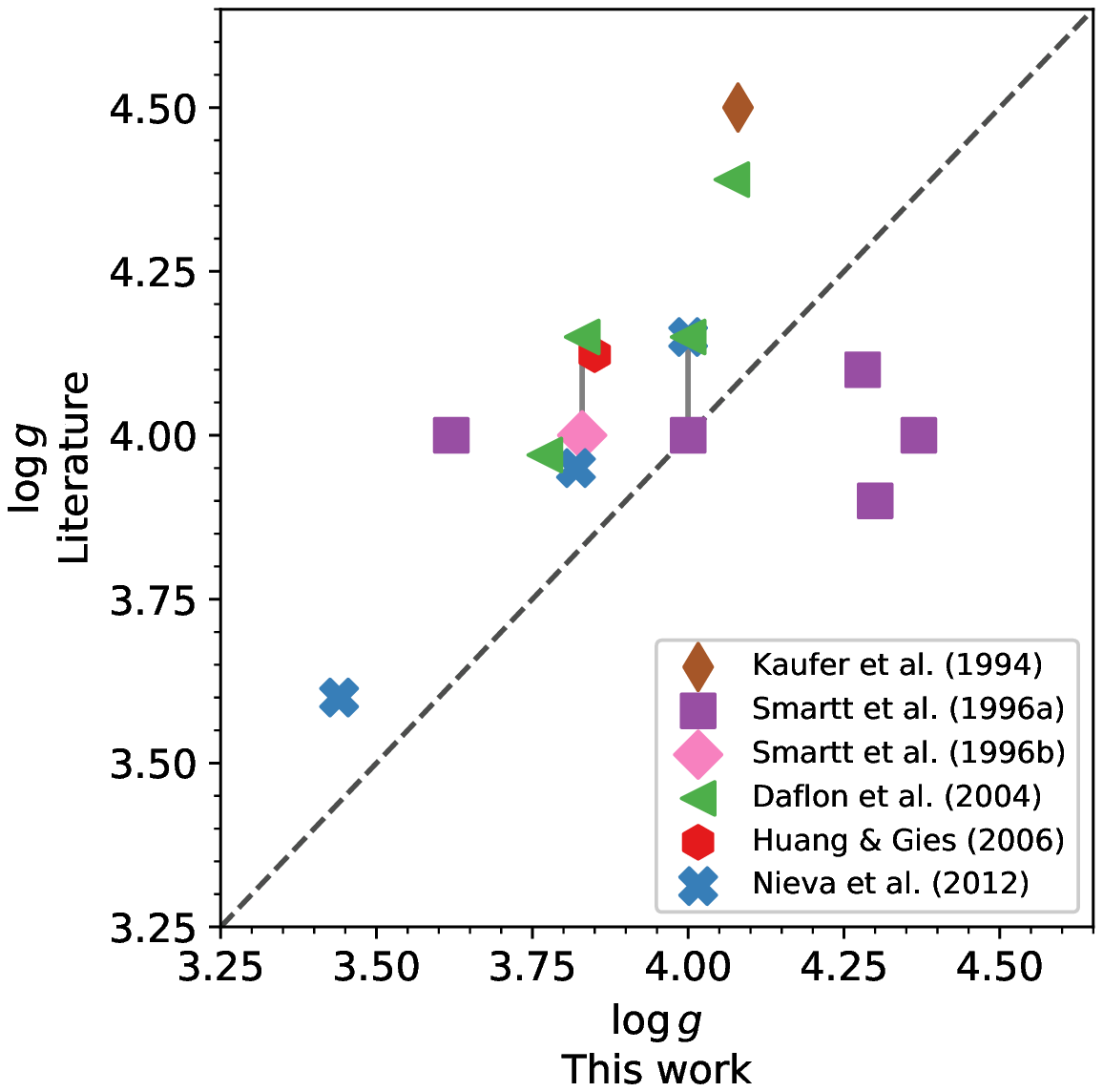}
\includegraphics{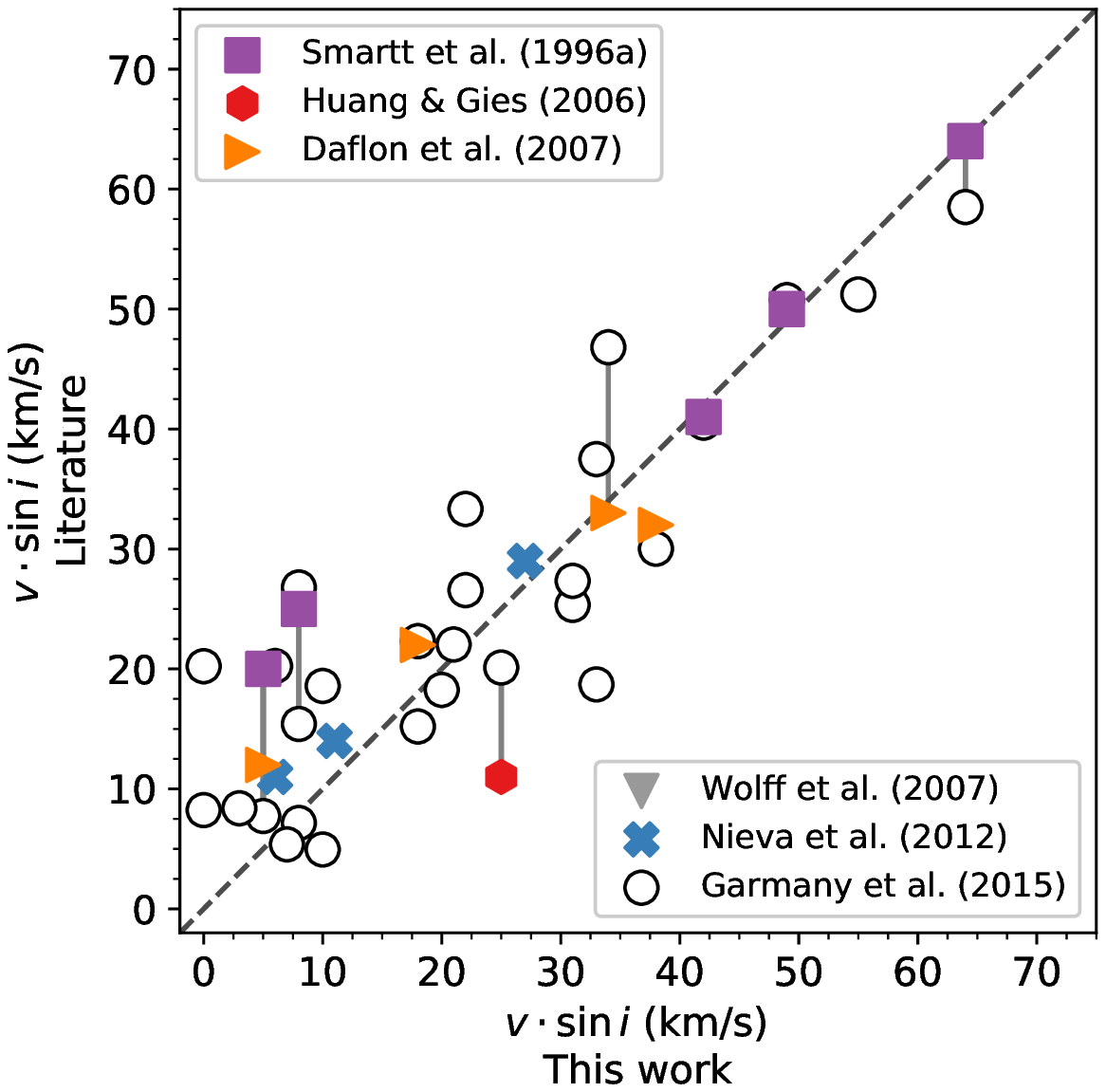}
}
\caption{
Comparison between effective temperature (left panel), surface gravity (middle panel), and projected rotational velocity (right panel) from Table~\ref{tab:sample_results} with values listed in previous studies in the literature, identified by different symbols.
}
\label{fig:comp}
\end{figure*}

\begin{figure*}
\centering
\resizebox{0.75\hsize}{!}{
\includegraphics{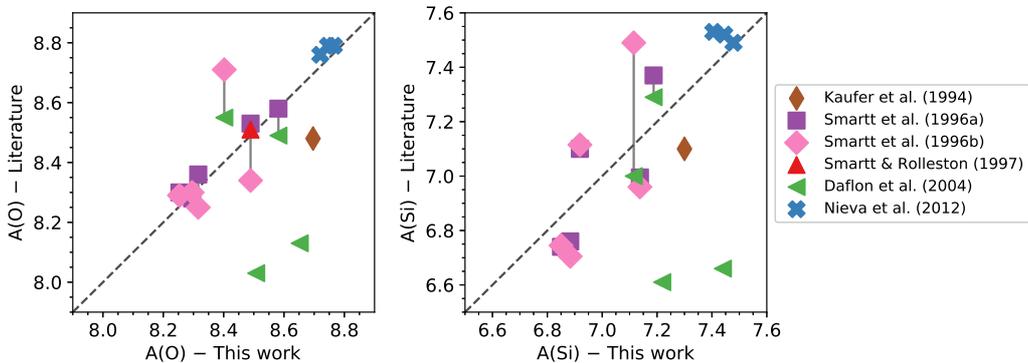}
}
\caption{ Comparison between O and Si abundances (left and right panel, respectively),  obtained in this work and in the literature for stars in common.
\label{fig:comp_abund} }
\end{figure*}

Several previous studies in the literature have measured the projected rotational velocities ($\vsini$) of Galactic OB stars. \cite{daflon2007}, \cite{huang2006}, \cite{wolff2007} derived $\vsini$ for large samples of OB stars from analyses of \ion{He}{i} lines and in some cases also \ion{Mg}{ii} lines.
\cite{garmany2015} also presented $\vsini$ values for the stars studied here based on measurements of full width at half maximum of three \ion{He}{i} lines interpolated in a grid of synthetic widths provided by \cite{daflon2007}. Previous studies of chemical abundances of OB stars include the earlier study by \cite{kaufer1994}, which  traced the radial abundance gradients based on LTE abundances for a sample of 16 OB stars located between 7 and 16\,kpc from the Galactic center. \cite{smartt1996a} analyzed six stars towards the Galactic anti-centre using line-blanketed LTE model atmospheres and spectral synthesis, and added two stars in a subsequent study \citep{smartt1996b} based on  a line-by-line differential abundance methodology. \cite{smartt1997} then compiled the abundance results of several previous works of their group, updated to account for non-LTE line formation calculations. Radial abundance gradients for seven elements have been studied in a sample of 69 OB stars by \cite{daflon2004}, based on non-LTE spectral synthesis.

A comparison of our results of  $\teff$, $\logg$, and $\vsini$ for stars found to be in common with the studies mentioned above are found in Fig.~\ref{fig:comp}. Our $\teff$-values show reasonable agreement to within $\sim$10\% with respect those effective temperatures that are based on spectroscopy: the mean difference $<$us$-$them$>=-320\pm1290$. Notable differences are found when comparing with the $\teff$ values from \cite{garmany2015}, which are based on a photometric calibration for the reddening-free index Q and therefore are expected to be more uncertain.  The  $\logg$ values show differences of up to 0.4\,dex  (middle plot of Fig.~\ref{fig:comp}), despite the fact that in all cited studies the gravities have been obtained by fitting the wings of H lines. The mean difference $<$us$-$them$>=-0.10\pm0.25$. We note that this offset is the same uncertainty adopted for the $\logg$ (see Sect.~\ref{sec:uncertainties}) and has minimal impact on the final abundances. The projected rotational velocities obtained in this study are overall in good agreement with the literature values (right panel in Fig.~\ref{fig:comp}), except for a slightly larger scatter for very sharp lined stars ($\vsini \lesssim 20 \,\kms$). The difference between our  $\vsini$ results and those listed in the sources presented in Fig.~\ref{fig:comp} is on average $-3.5\pm8.7 \kms$. The lower $\vsini$ values we report may be the result of including the  macroturbulent velocity as an extra parameter of line broadening in our spectrum synthesis analysis.

Comparisons between our derived abundances of oxygen and silicon and the literature values are shown in Fig.~\ref{fig:comp_abund}. We note, however, that for two stars (ALS~404 and ALS~428) there is a larger discrepancy in the two sets of results. The large discrepancy in the oxygen and silicon abundances for these two stars may be due to differences in the adopted line lists,  $\teff$, and/or the microturbulence. In particular, the effective temperatures in \cite{daflon2004} are photometric, based on a calibration for the reddening-free indice Q. The resolutions of the spectra analyzed in the two studies are comparable, although the data of the present study has higher signal-to-noise ratio (we used a larger telescope), the equivalent widths measured are consistent between the two studies.


\subsection{Stellar parameters and abundance results}
\label{sec:results}

\begin{figure}
\centering
\resizebox{\hsize}{!}{
\includegraphics{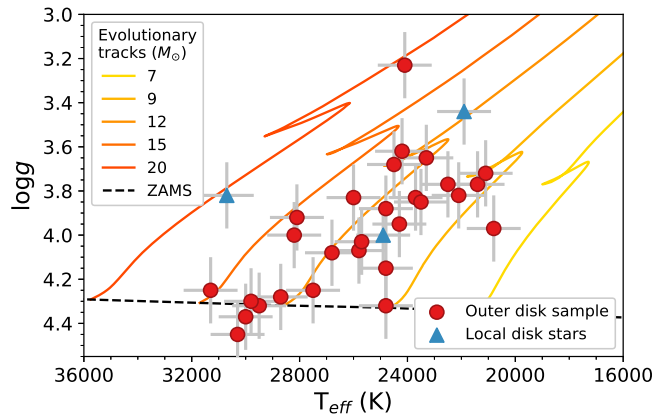}}
\caption{The spectroscopic HR diagram for the sample stars: red circles	represent the  outer disk stars and  the blue triangles represent the	Local disk stars. Evolutionary tracks computed for	masses  7,9, 12, 15, and 20 $M_{\sun}$ (solid lines, from yellow to orange, respectively) and for the Zero Age Main	Sequence (ZAMS, dashed line)  are also shown \citep{ekstrom2012}.
\label{fig:teff_logg}}
\end{figure}

\begin{figure*}
\centering
\resizebox{0.9\hsize}{!}{
\includegraphics{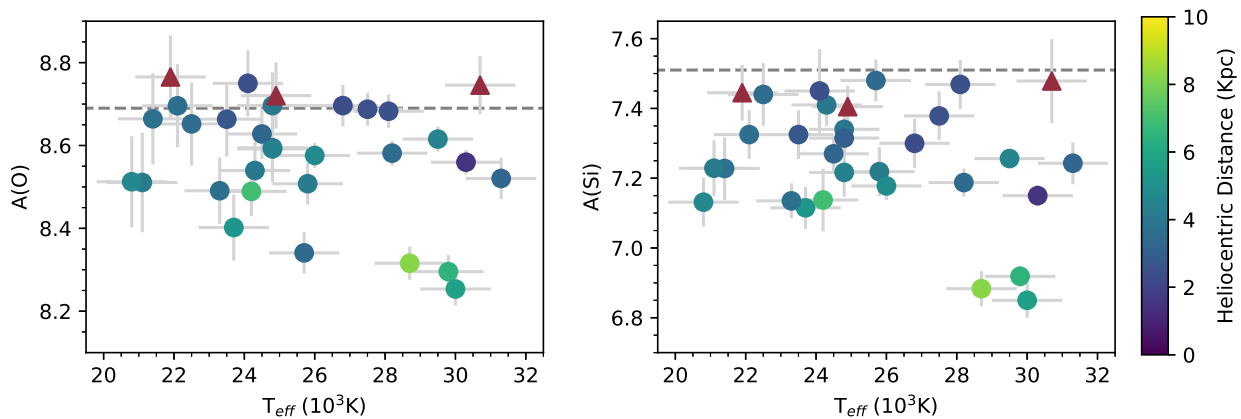}}
\caption{Silicon and Oxygen abundances as a function of $\teff$.  Circles	represent the stars of the outer disk sample, color-coded by distance; triangles represent	the three nearby stars.  The dashed lines in both panels	represent the solar abundances $A(\rm{O})=8.69$ and $A(\rm{Si}) = 7.51$	\citep{asplund2009}.  A slight, apparent correlation between	abundances and $\teff$ is likely connected to the radial gradient.
	\label{fig:teff_abund}
 }
\end{figure*}

Figure~\ref{fig:teff_logg} shows the studied sample of 28 early B-type outer disk stars, as well as the three solar neighborhood stars, in the $ \teff-\logg$ plane, with evolutionary tracks from \cite{ekstrom2012} calculated for non-rotating stars at $Z=0.014$ and colour-coded according to the initial mass,  ranging from 7 to 20\,M$_\odot$.
Almost all studied stars have masses in the range 7 to 15\,M$_\odot$ and fall between the zero-age main sequence (ZAMS) and the terminal-age main sequence.
ALS~921 is the most evolved star of our sample, with $\logg=3.23$, $M_{\star}\sim 20\,M_\odot$ and a high microturbulence value of $\vmic=10 \kms$.

Our iterative procedure based on spectral synthesis treats the broadening mechanisms (projected rotational velocity $\vsini$, microturbulent velocity $\vmic$, and radial-tangential macroturbulent velocity $\vmac$) as if they were independent.
$\vmic$-values for the studied stars span from 2 to 14 $\kms$, with most of the stars (27 stars) with $\vmic \leq 8 \kms$ and an average value of 4.6$\pm$1.5 $\kms$. Two stars with $\vmic$ = 9 and 10 $\kms$ are the most evolved stars of our sample, respectively $\logg$ = 3.44 and 3.23, which  is consistent with generally higher microturbulences associated with lower gravities. On the other hand, stars ALS~18020 and HD~63922 show $\vmic$ higher than expected for their $\logg$ values. We note, however, that lower $\vmic$ for these stars would produce exceptionally high abundances. As discussed in Section \ref{sec:check} for HD~63922, one of the stars used to test our methodology, the final abundances are consistent with previous values from the literature, despite of some differences in the line broadening parameters.
The $\vmac$ and $\vsini$ parameters broaden spectral lines in very similar ways which makes it very difficult to disentangle the effects of these parameters in those spectra with low $S/N$ ratios. The $\vsini$ broadens the full profile as the equivalent width remains constant, while the macroturbulent velocity affects mainly the wings of the metallic lines.
We note that for spectra with lower $S/N$ ratios, when the line wings are not well defined, the $\vmac$-values tend to be less accurate. The $\vmac$ values also tend to be lower than $20\,\kms$, with only a few stars reaching values up to about $50\,\kms$, which are typical values for massive OB stars \citep{simondiaz2017}.
We stress that the sample is biased to slow rotators by design, that is, most stars of our sample have $\vsini < 50\,\kms$ to allow for a proper detailed elemental abundance study.

The non-LTE abundances of O and Si as a function of the $\teff$
are presented in Fig.~\ref{fig:teff_abund}.  The dashed lines in both panels represent the solar abundances $A(\rm{O})=8.69$ and $A(\rm{Si}) = 7.51$ \citep{asplund2009}. The outer disk stars are represented by circles, color-coded according to the distance. Figure~~\ref{fig:teff_abund} also includes the abundances derived for the three OB stars in the Solar neighborhood (shown as triangles).
The three stars with lowest O and Si abundances and $\teff$ around 28\,000-30\,000 K are located at larger distances, indicating that the slight, apparent trend is probably a consequence of the radial abundance gradient.
The lack of a correlation between abundances and $\teff$ suggests that the methodology is consistent throughout the entire effective temperature range.


\section{Distances}
\label{sec:distances}

One of the goals in this work is to obtain the radial distribution of O and Si abundances in the Galactic disk.  The heliocentric distances (d$_{\sun}$), a key parameter for such an analysis, were unknown  a priori for most of the sample  stars, although the star's location towards the
outer Galactic disk was one important selection criterion of the observed sample (Sect.~\ref{sec:observation}).

In \cite{garmany2015}, we obtained heliocentric distances  for the full observed sample  (including the 28 stars studied here) based on  intrinsic colors and a calibration between absolute magnitude and spectral type. We estimated the uncertainties of these heliocentric distances assuming a typical error of $\sim$0.06\,dex in the apparent magnitude (column~6 of Table~\ref{tab:distances}). The second Data Release \citep{gaiadr2} provides state-of-the-art measurements of parallaxes for 31 stars analyzed in this work, including the three stars to represent the  Local disk.
For our objects, the errors on the parallaxes are 21\,\%, on average, with one star having a relative error of 103\,\% (ALS~18681). \cite{bailerjones2018}, using Bayesian statistics and a pure geometric prior, estimated the distances for approximately 1.33 billion stars of the second Gaia
Data Release, and we present their distances for the full sample stars in column~7 of Table~\ref{tab:distances}.

We can estimate the distances to these stars more precisely than using \emph{just} the Gaia parallaxes \citep[as done by][]{bailerjones2018} if we use further information about the star. The stellar parameters and apparent colour of the star (in combination with theoretical models) tell us about the intrinsic luminosity of the star, and the extinction towards it. Comparing this to the apparent magnitude tells us the distance. Our approach incorporates all of this information (including the Gaia parallax) with the associated uncertainties in a Bayesian framework. For stars which have small relative errors on their Gaia parallax, our results are very similar to those from using the parallax alone. When the relative errors on the Gaia parallax are large, the additional information we use allow us to make much more accurate distance estimates.

 Specifically, our estimates are derived in the manner 
described in \cite{mcmillan2018}, which follows the pioneering approach of \cite{burnett2010}. This scheme uses a prior and the observational data for a star to estimate the probability density function in the parameters of initial mass ${\cal M}$, age $\tau$, metallicity $[\mathrm{M}/\mathrm{H}]$, distance $s$ and  line-of-sight extinction (parametrized as $A_V$) for the star. Isochrones from \cite{bressan2012} are used to give the expected observables for these parameters, and to give distance estimates we marginalize over all the other parameters.

\begin{figure}
\centering
\resizebox{\hsize}{!}{
\includegraphics{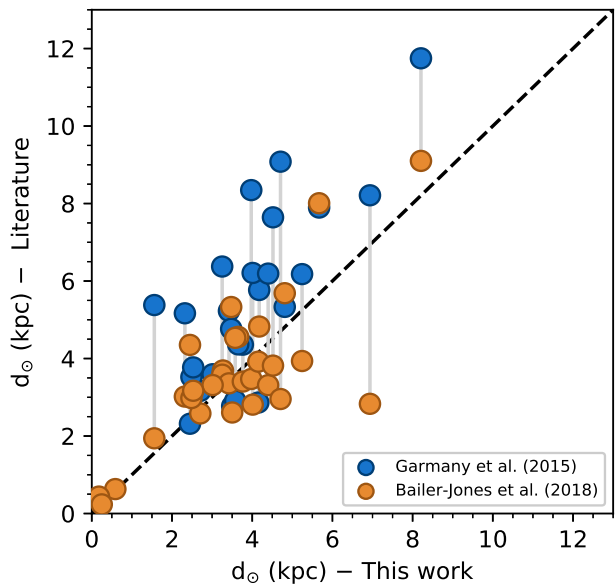}
}
\caption{Comparison between the heliocentric distances obtained in this work	versus \cite{garmany2015} (blue circles) and \cite{bailerjones2018} 	(yellow circles).
	 }
\label{fig:dist_comparison}
\end{figure}
\begin{figure}
\centering
\resizebox{\hsize}{!}{
\includegraphics{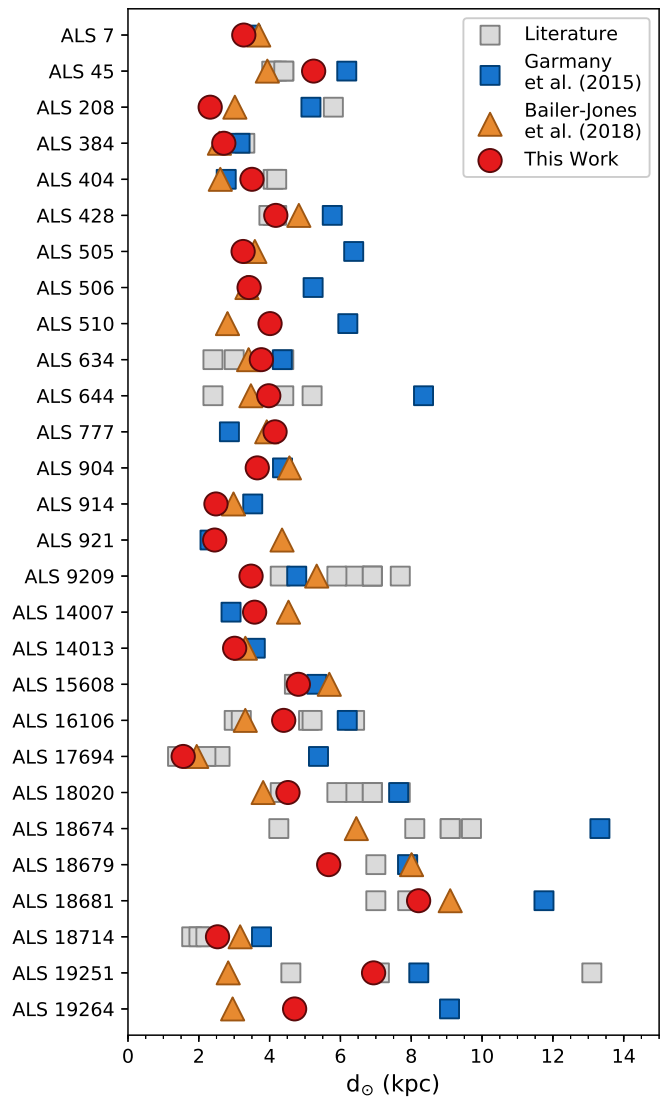}}
\caption{Comparison of heliocentric distances star by star for results	calculated in this work (red circles), \cite{bailerjones2018} (yellow 	triangles), by \cite{garmany2015} (blue squares) and obtained in the	literature (gray squares).
	}
\label{fig:distances_strip}
\end{figure}

In this case we use the ``Density" prior described by \cite{mcmillan2018},  which uses a dust extinction prior based on that from \cite{sharma2011}, and a flat prior on metallicity and age. The observables we use are 2MASS $J$, $H$
and $K_s$ magnitudes, $\teff$, $\log g$, and A(O) from this study (where we make the approximation A(O)$-8.69\approx[\mathrm{M}/\mathrm{H}]$), and {\it Gaia} DR2 parallaxes. We make an effort to correct the {\rm Gaia} parallaxes for a zero-point offset and systematic uncertainties. We apply a global parallax zero-point correction of $29\;\mu\,\mathrm{as}$, and we add an uncertainty of $43\;\mu\,\mathrm{as}$ in quadrature with the quoted errors,
corresponding to the values estimated  by \cite{lindegren2018}. The  distances obtained are listed in column 8 of Table \ref{tab:distances}.

\begin{sidewaystable*}
\centering
\caption{Galactic coordinates and distances for the sample
	stars.
\label{tab:distances}
}
\footnotesize
\setlength{\tabcolsep}{0.5mm}
\begin{tabular}{@{\extracolsep{4pt}}llcclccccc@{}}
\hline\hline
\noalign{\smallskip}
    Star                                            &
    Gaia DR 2 ID                                    &
    l ($^o$)                                        &
    b ($^o$)                                        &
    Cluster or                                      &
    \multicolumn{4}{c}{Heliocentric distance (kpc)} &
    Galactocentric \\
    \cline{6-9}
    \noalign{\smallskip}
                                                    &
                                                    &
                                                    &
                                                    &
   HII region                                       &
   \cite{garmany2015}                               &
   \cite{bailerjones2018}                           &
   This work                                        &
    Literature                                      &
   distance (kpc) \\
   (1) & (2) & (3) & (4) & (5) & (6) & (7) & (8) & (9) & (10) \\
\hline
\noalign{\smallskip}
\multicolumn{7}{l}{Outer disk sample stars} \\
\noalign{\smallskip}
\hline
\noalign{\smallskip}
\noalign{\smallskip}
 ALS 7     & 3398994894234581376 & 188.8 & -2.6           & --                     & \phantom{1}3.5$\pm$0.2 & 3.69$\pm$0.56 & 3.27$\pm$0.34 & --                                                                                            & 11.57$\pm$0.49            \\
 ALS 45    & 3372477521332283008 & 192.4 & \phantom{-}3.3 & Cluster: Bochum 1      & \phantom{1}6.2$\pm$0.4 & 3.94$\pm$0.81 & 5.24$\pm$0.95 & 4.1\uprefs{1}, 4.2\uprefs{25}, 4.4\uprefs{4, 20}                                              & 13.49$\pm$1.00            \\
 ALS 208   & 2934266685350452480 & 231.4 & -4.4           & HII region: SH 2-301   & \phantom{1}8.1$\pm$0.4 & 3.02$\pm$0.41 & 2.32$\pm$0.04 & 5.8\uprefs{4}, 11.0\uprefs{36}                                                                & \phantom{1}9.94$\pm$0.35  \\
 ALS 384   & 5619910010422265216 & 235.4 & -2.4           & Cluster: NGC 2384      & \phantom{1}3.2$\pm$0.2 & 2.58$\pm$0.29 & 2.71$\pm$0.32 & 1.4\uprefs{36}, 3.3\uprefs{7}                                                                 & 10.11$\pm$0.41            \\
 ALS 404   & 3028413158753538432 & 231.1 & \phantom{-}0.2 & --                     & \phantom{1}2.8$\pm$0.2 & 2.61$\pm$0.23 & 3.50$\pm$1.00 & 2.4\uprefs{36}, 4.1\uprefs{20}, 4.2\uprefs{5}                                                 & 10.88$\pm$0.87            \\
 ALS 428   & 3028602309115840640 & 230.7 & \phantom{-}1.0 & --                     & \phantom{1}5.8$\pm$0.4 & 4.83$\pm$0.85 & 4.17$\pm$0.44 & 3.7\uprefs{36}, 4.1\uprefs{20}, 4.2\uprefs{5}                                                 & 11.44$\pm$0.50            \\
 ALS 505   & 3028360244759075712 & 231.4 & \phantom{-}1.7 & --                     & \phantom{1}6.4$\pm$0.5 & 3.58$\pm$0.56 & 3.25$\pm$0.17 & 0.5\uprefs{36}                                                                                & 10.66$\pm$0.37            \\
 ALS 506   & 3028381788314791168 & 231.2 & \phantom{-}1.9 & --                     & \phantom{1}5.2$\pm$0.4 & 3.36$\pm$0.51 & 3.42$\pm$0.36 & --                                                                                            & 10.80$\pm$0.44            \\
 ALS 510   & 3028382265045578624 & 231.2 & \phantom{-}2.0 & --                     & \phantom{1}6.2$\pm$0.4 & 2.81$\pm$0.39 & 4.01$\pm$0.62 & --                                                                                            & 11.28$\pm$0.61            \\
 ALS 634   & 5588928001826043648 & 247.8 & -5.5           & Cluster: Bochum 15     & \phantom{1}4.4$\pm$0.3 & 3.41$\pm$0.29 & 3.77$\pm$0.54 & 2.4\uprefs{32}, 3.0\uprefs{3},4.4\uprefs{2}, 7.5\uprefs{36}                                   & 10.35$\pm$0.48            \\
 ALS 644   & 5588897219803862016 & 248.2 & -5.7           & Cluster: Bochum 15     & \phantom{1}8.3$\pm$0.6 & 3.47$\pm$0.43 & 3.98$\pm$0.47 & 2.4\uprefs{32}, 4.4\uprefs{2}, 5.2\uprefs{3}                                                  & 10.47$\pm$0.45            \\
 ALS 777   & 5598644901475541888 & 246.1 & -2.8           & --                     & \phantom{1}2.9$\pm$0.2 & 3.92$\pm$0.43 & 4.15$\pm$0.69 & 2.5\uprefs{36}                                                                                & 10.70$\pm$0.59            \\
 ALS 904   & 5601220821009448320 & 244.5 & \phantom{-}1.3 & --                     & \phantom{1}4.4$\pm$0.3 & 4.56$\pm$0.78 & 3.65$\pm$0.50 & --                                                                                            & 10.44$\pm$0.48            \\
 ALS 914   & 5597822436730855040 & 245.6 & \phantom{-}0.8 & --                     & \phantom{1}3.5$\pm$0.3 & 2.98$\pm$0.46 & 2.48$\pm$0.24 & 4.8\uprefs{36}                                                                                & \phantom{1}9.63$\pm$0.37  \\
 ALS 921   & 5594504522951919744 & 248.9 & -1.2           & --                     & \phantom{1}2.3$\pm$0.2 & 4.35$\pm$0.63 & 2.45$\pm$0.12 & 1.7\uprefs{36}                                                                                & \phantom{1}9.49$\pm$0.35  \\
 ALS 9209  & 3112494908868923520 & 213.9 & \phantom{-}0.6 & HII region: SH 2-285   & \phantom{1}4.8$\pm$0.3 & 5.33$\pm$1.29 & 3.48$\pm$0.30 & 2.1\uprefs{36}, 4.3\uprefs{13}, 5.9\uprefs{20}, 6.4\uprefs{8}, 6.9\uprefs{22}, 7.7\uprefs{23} & 11.38$\pm$0.44            \\
 ALS 14007 & 2929941275327270400 & 234.1 & -3.8           & --                     & \phantom{1}2.9$\pm$0.2 & 4.53$\pm$0.79 & 3.58$\pm$0.46 & 3.2\uprefs{36}                                                                                & 10.82$\pm$0.49            \\
 ALS 14013 & 2930437636108728064 & 234.1 & -2.7           & --                     & \phantom{1}3.6$\pm$0.3 & 3.32$\pm$0.55 & 3.01$\pm$0.35 & --                                                                                            & 10.38$\pm$0.43            \\
 ALS 15608 & 5597787595954892416 & 245.8 & \phantom{-}0.5 & Cluster: Ruprecht 44   & \phantom{1}5.3$\pm$0.4 & 5.69$\pm$0.94 & 4.81$\pm$0.57 & 2.5\uprefs{36}, 4.7\uprefs{32}                                                                & 11.20$\pm$0.53            \\
 ALS 16106 & 5602035005068474880 & 243.1 & \phantom{-}0.5 & Cluster: Haffner 19    & \phantom{1}6.2$\pm$0.4 & 3.31$\pm$0.54 & 4.40$\pm$0.58 & 3.0\uprefs{34}, 3.2\uprefs{34}, 5.1\uprefs{28}, 5.2\uprefs{18}, 6.4\uprefs{29}                & 11.04$\pm$0.54            \\
 ALS 17694 & 3375202527758417664 & 190.1 & \phantom{-}0.5 & HII region: SH 2-252 E & \phantom{1}5.4$\pm$0.4 & 1.94$\pm$0.28 & 1.56$\pm$0.08 & 1.4\uprefs{30}, 2.2\uprefs{22}, 2.6\uprefs{26}                                                & \phantom{1}9.87$\pm$0.36  \\
 ALS 18020 & 3112496794360485248 & 213.8 & \phantom{-}0.6 & HII region: SH 2-285   & \phantom{1}7.6$\pm$0.5 & 3.82$\pm$0.65 & 4.52$\pm$0.26 & 4.3\uprefs{13}, 5.9\uprefs{20}, 6.4\uprefs{8}, 6.9\uprefs{22}, 7.7\uprefs{23}                 & 12.34$\pm$0.42            \\
 ALS 18674 & 3119827723711464576 & 210.8 & -2.5           & HII region: SH 2-283   & 13.3$\pm$1.0           & 6.45$\pm$1.16 & --            & 4.3\uprefs{4}, 8.1\uprefs{23}, 9.1\uprefs{10,22}, 9.7\uprefs{14}                              & 14.25$\pm$1.16            \\
 ALS 18679 & 3099408968152006272 & 218.8 & -4.6           & HII region: SH 2-289   & \phantom{1}7.9$\pm$0.6 & 8.01$\pm$1.77 & 5.67$\pm$0.59 & 7.0\uprefs{14}, 7.9\uprefs{4}                                                                 & 13.22$\pm$0.63            \\
 ALS 18681 & 3099397148402008192 & 218.8 & -4.6           & HII region: SH 2-289   & 11.7$\pm$0.8           & 9.10$\pm$2.16 & 8.21$\pm$0.71 & 7.0\uprefs{14}, 7.9\uprefs{4}                                                                 & 15.57$\pm$0.74            \\
 ALS 18714 & 5619746732941651584 & 235.6 & -3.8           & Cluster: NGC 2367      & \phantom{1}3.8$\pm$0.3 & 3.17$\pm$0.50 & 2.53$\pm$0.23 & 1.2\uprefs{36}, 1.8\uprefs{32}, 2.0\uprefs{21}, 2.2\uprefs{27}                                & \phantom{1}9.98$\pm$0.38  \\
 ALS 19251 & 3429605538471099008 & 184.7 & \phantom{-}1.1 & --                     & \phantom{1}8.2$\pm$0.6 & 2.83$\pm$0.54 & 6.94$\pm$1.06 & 4.6\uprefs{14}, 7.1\uprefs{9}, 13.1\uprefs{17}                                                & 15.25$\pm$1.12            \\
 ALS 19264 & 3432289583791213696 & 185.7 & \phantom{-}5.9 & --                     & \phantom{1}9.1$\pm$0.6 & 2.96$\pm$0.43 & 4.71$\pm$0.75 & --                                                                                            & 13.00$\pm$0.82            \\
\noalign{\smallskip}
\hline
\noalign{\smallskip}
\multicolumn{7}{l}{Local disk sample stars} \\
\noalign{\smallskip}
\hline
\noalign{\smallskip}
 HD 61068  & 5716409335624704768 & 235.5 & \phantom{-}0.6 & --                     & --                     & 0.63$\pm$0.04 & 0.59$\pm$0.03 & 0.4\uprefs{31}, 0.5\uprefs{15,24,33}                                                          & \phantom{1}8.68 $\pm$0.35 \\
 HD 63922  & 5530670107648495104 & 260.2 & -10.2          & --                     & --                     & 0.44$\pm$0.14 & 0.18$\pm$0.01 & 0.4\uprefs{31}, 0.5\uprefs{15}, 0.6\uprefs{24}                                                & \phantom{1}8.36 $\pm$0.35 \\
 HD 74575  & 5639188675500905216 & 255.0 & \phantom{-}5.8 & --                     & --                     & 0.24$\pm$0.02 & 0.25$\pm$0.02 & 0.3\uprefs{15, 24, 31}                                                                        & \phantom{1}8.40 $\pm$0.35 \\
\noalign{\smallskip}
\hline
\end{tabular}

\tablebib{

(1)~\cite{moffat1975},
(2)~\cite{fitzgeraldhurkens1976},
(3)~\cite{jackson1977},
(4)~\cite{Moffat1979},
(5)~\cite{fitzgeraldmoffat1980},
(6)~\cite{Blitz1982},
(7)~\cite{babu1983},
(8)~\cite{lahulla1987},
(9)~\cite{philip1990},
(10)~\cite{fich1991b},
(11)~\cite{fitzsimmone1993},
(12)~\cite{turbide1993},
(13)~\cite{rolleston1994},
(14)~\cite{smartt1996a},
(15)~\cite{perryman1997},
(16)~\cite{Hensberge2000},
(17)~\cite{rolleston2000},
(18)~\cite{morenocorral2002},
(19)~\cite{park2002},
(20)~\cite{daflon2004},
(21)~\cite{kharchenko2005},
(22)~\cite{rudolph2006},
(23)~\cite{Russeil2007},
(24)~\cite{vanleeuwen2007},
(25)~\cite{Bica2008},
(26)~\cite{reid2009},
(27)~\cite{melnik2009},
(28)~\cite{pandey2010},
(29)~\cite{Vazquez2010},
(30)~\cite{bonatto2011},
(31)~\cite{nieva2012},
(32)~\cite{Kharchenko2013},
(33)~\cite{lyumbikov2013},
(34)~\cite{perren2015},
(35)~\cite{garmany2015},
(36)~\cite{gaiadr1},
}
\end{sidewaystable*}

Figure~\ref{fig:dist_comparison} shows a comparison between the distances obtained in this work  and the distances published  by \cite{garmany2015} (blue circles) and by \cite{bailerjones2018} (yellow circles), and Fig.~\ref{fig:distances_strip} shows an alternative view of this comparison between different distances for each star of our sample. Different points in each line of this figure represent the distinct distance values in the literature (gray
squares) as well as the distance obtained in this work (red circles), by \citet{garmany2015} (blue squares) and by \cite{bailerjones2018} (yellow triangles). The stars are identified by the ALS number in the y-axis.
From Figures \ref{fig:dist_comparison} and \ref{fig:distances_strip}, we note that the distances we derived and those obtained by \cite{bailerjones2018} are in
good agreement, while the distances obtained by \cite{garmany2015} purely from photometric parameter are 
overestimated for most of the stars.

The Galactocentric distances projected onto the Galactic plane were computed using the Galactocentric distance of the Sun
$R_{\odot}=8.33\pm0.35$ kpc from \cite{gillessen2009}.
It was not possible to calculate the distance for ALS~18674 due to lack of 2MASS JHK photometry, hence we used the heliocentric
distance calculated by \cite{bailerjones2018} to obtain the Galactocentric distance.  The calculated Galactocentric distances for the sample stars are listed in column 10
of Table~\ref{tab:distances}.

\section{Radial abundance gradients}
\label{sec:gradients}

\begin{figure*}
\centering
\resizebox{0.9\hsize}{!}{
\includegraphics{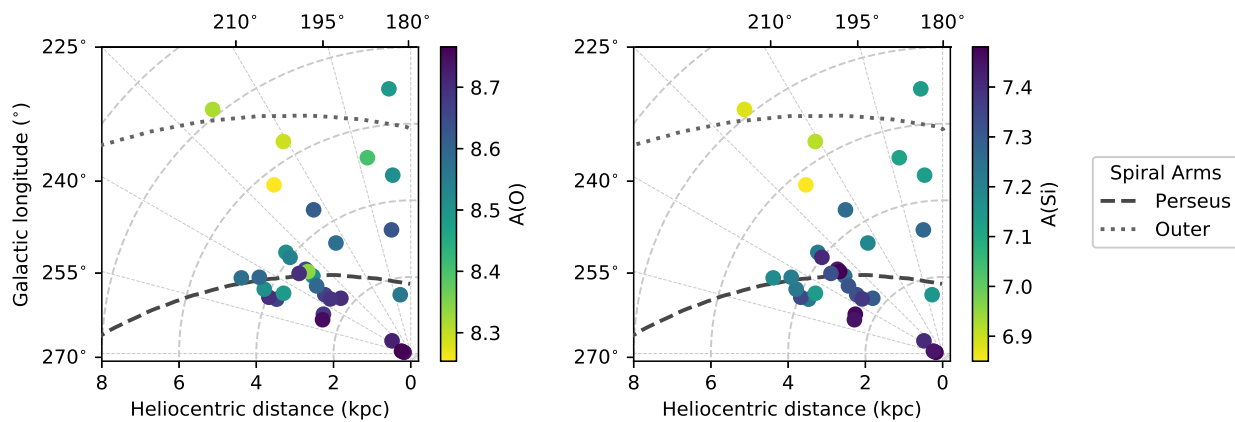}}
\caption{Distribution of chemical abundances of oxygen (left) and silicon	(right) projected onto the Galactic plane. The location of the Perseus	Arm (dashed curve) and the Outer Arm (dotted curve), from 	\cite{carraro2015}, is also shown. A radial gradient is visible on both plots while an azimuthal gradient is not so clear. \label{fig:azi_grad} }
\end{figure*}

Massive OB stars represent the present abundance distribution in the Galactic disk and radial abundance gradients traced by OB stars can help put observational constraints on Galactic chemical evolution models.
Figure~\ref{fig:azi_grad} shows the distribution of the oxygen (left panel) and silicon (right panel) abundances projected onto the Galactic plane. The abundances are color-coded from dark blue (highest abundances) to light yellow (lowest abundances). All stars of our sample are located in the third Galactic quadrant and the three stars near (0,0) in longitude and latitude represent the solar neighborhood stars HD~61068, HD~63922, and HD~74575. The Galactic spiral arms proposed by \cite{carraro2015} are overplotted: the Perseus Arm (dashed curve), which is located near to the Sun and the stars that could be associated with this arm have heliocentric distances from 1 to 5\,kpc, and the Outer Arm (dotted curve) which traverses the third quadrant have heliocentric distances from 6 to 10\,kpc. From the figure, we note that a radial gradient does exist whereas the existence of an azimuthal gradient is not clear.

\begin{figure*}
\centering
\resizebox{0.9\hsize}{!}{
\includegraphics{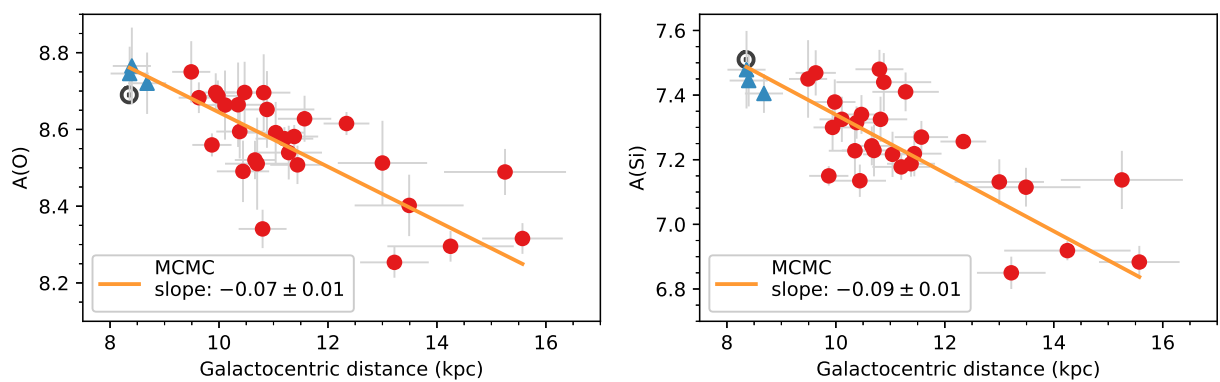}}
\caption{Abundances of O and Si for the Outer disk sample stars (circles) and 	Local disk stars (triangles) as a function of the Galactocentric 	distance. The Sun ($\odot$) is represented at $R_G = 8.33$ Kpc for 	reference.  A Markov Chain Monte Carlo (MCMC) linear regression shows 	the abundance gradients for the distance covered by the sample stars.
	\label{fig:rad_grad}
}
\end{figure*}

The distribution of the stars' abundances according to their projected Galactocentric distances is shown in Fig.~\ref{fig:rad_grad}, where the left and right panels show the results for oxygen and silicon, respectively, for the outer disk stars (red circles) and nearby stars (blue triangles).
The Sun is represented in the figure with $A({\rm O})=8.69\pm0.05$ and $A({\rm Si})=7.51\pm0.03$ \citep{asplund2009} and  $R_G = 8.33$\,kpc
\citep{gillessen2009}.  On both panels, the distribution follows the usual pattern of a negative abundance gradient: as the Galactocentric distance increases, the abundance decreases (see Table \ref{tab:rad_grad_summary}).

To derive the radial abundance gradients for our sample, we used a Markov Chain Monte Carlo (MCMC) linear regression algorithm that was built according to \cite{foremanmackey2013, hogg2010, astroML}. The MCMC approach is more appropriate in this analysis because: (i) it is less sensitive to outliers when compared to the standard linear fitting method, (ii) it considers the errors on both variables, in this case abundances and distances, and (iii) it is suitable when the uncertainties may be high (which is the case of some of the distances).
The  radial abundance gradients obtained for $8.4 \leq R_G \leq 15.6 \enspace{\rm kpc}$, are:

\begin{equation}
    \begin{array}{rcl}
        {\rm A(O)}  & = & 9.36\pm0.10 - 0.07\pm0.01 \cdot R_G, \\
        {\rm A(Si)} & = & 8.30\pm0.12 - 0.09\pm0.01 \cdot R_G.
    \end{array}
\end{equation}

The sample of OB stars studied has been selected in order to  cover the outer Galactic disk. The radial gradients we derived predict O and Si abundances at $R_G\sim 8.5$\,kpc that are fully consistent with the abundances of the three nearby stars and the Sun that represent the abundances in the solar
neighborhood.

\begin{table*}
\centering
\caption{Radial gradient slopes.
	\label{tab:rad_grad_summary}
}
\footnotesize
\begin{tabular}{lccccc}
\hline\hline
\noalign{\smallskip}
\multirow{2}{*}{}                    &
\multicolumn{2}{c}{Slope (dex/kpc)}  &
\multirow{2}{*}{Range (kpc)}         &
Type of                              \\
                 &
Oxygen           &
Silicon          &
                 &
Objects          \\
\noalign{\smallskip}
\hline
\noalign{\smallskip}
This work              & $-0.07\pm0.01$          & $-0.09\pm0.01$ & \phantom{1}8.4 -- 15.6 & OB Stars \\
\cite{gummersbach1998} & $-0.067\pm0.024$          & $-0.107\pm0.028$ & \phantom{1}5.0 -- 14.0 & OB Stars \\
\cite{rolleston2000}   & $-0.067\pm0.008$          & $-0.06\pm0.01$ & \phantom{1}6.0 -- 18.0 & OB Stars \\
\cite{daflon2004}      & $-0.031\pm0.012$          & $-0.040\pm0.017$ & \phantom{1}4.7 -- 13.2 & OB Stars \\
\cite{lemasle2013}     & --                & $-0.057\pm0.011$ & \phantom{1}7.0 -- 15.0 & Cepheids \\
\cite{genovali2015}    & --                & $-0.049\pm0.002$ & \phantom{1}4.0 -- 18.5 & Cepheids \\
\cite{luck2018}    &  $-0.0429\pm0.0023$ & $-0.0497\pm0.0022$ & $\sim$ 4.5 -- 17.0 & Cepheids \\
\cite{esteban2017}     & $-0.040\pm0.005$          & --       & \phantom{1}5.1 -- 17.0 & HII regions \\
\cite{fernandez2017}   & $-0.053\pm0.009$ & --       & 11.0 -- 18.0           & HII regions \\
\cite{esteban-garcia2018} & $-0.051\pm0.008$       & --       & \phantom{1}7.0 -- 17.0 & HII regions \\
\noalign{\smallskip}
\hline
\end{tabular}
\end{table*}

In Table~\ref{tab:rad_grad_summary} we compile some results for the oxygen and silicon radial abundance gradients for young stellar populations in the Milky Way; for OB stars \citep{gummersbach1998,rolleston2000,daflon2004}, Cepheids
\citep{lemasle2013, genovali2015,luck2018}, and \ion{H}{ii} regions \citep{esteban2017,fernandez2017,esteban-garcia2018}.

 The radial gradient obtained in this study for silicon of $-0.09$ dex/Kpc is relatively steep.
Overall, the silicon gradients in OB stars from the literature span a relatively large range of values from flat, $-0.04$ dex/kpc, to quite steep, $-0.11$ dex/kpc; this somewhat large variation in the slopes in the different studies may be due to the impact of the microturbulent velocity on the silicon abundance determinations, which, in general, are based on intermediate to strong \ion{Si}{iii} lines. The silicon results for the Cepheids in the three studies presented in Table~\ref{tab:rad_grad_summary} (from \cite{lemasle2013},  \cite{genovali2015}, and \cite{luck2018}) show more consistency and indicate a mean slope of $-0.053$ dex/kpc.

The oxygen abundance gradients in the different OB star studies show a smaller variation in the derived slopes (from $-0.03$ to $-0.07$ dex/kpc) when compared to silicon.
The oxygen gradient obtained in this study for the OB stars of $-0.07$ dex/kpc, as well as two earlier studies from the literature (\cite{gummersbach1998} and \cite{rolleston2000}), are steeper than those for  Cepheids \citep{luck2018} and for \ion{H}{ii} regions presented in the three studies in Table~\ref{tab:rad_grad_summary}.

It is important to keep in mind, however, that the radial abundance gradients in the different studies were obtained for different galactocentric distance ranges and that the abundance gradients may not be uniform along the entire Galactic disk.
Results for the galactic open clusters, for example, seem to indicate that there is a break in the metallicity slope at Rg $\sim$ 12 kpc with the gradient being flatter for $R_G>12$ kpc   \citep{netopil2016,reddy2016,frinchaboy2013,cunha2016}. 

The possibility of external flattening of the oxygen gradient has been discussed by \cite{esteban2017}.
The authors divided their \ion{H}{ii} region sample into 'inner' and 'outer' disk samples (at $R_G=11.5$\,kpc) and obtained oxygen gradients for the inner and outer disk that are equal within the uncertainties, finding no evidence of flattening of the radial gradient to within $R_G\approx 17$\,kpc. More recently, \cite{esteban-garcia2018}
obtained a two-zone oxygen abundance gradient for his sample of \ion{H}{ii} regions finding that the oxygen slope is flatter in the inner disk, with a break at $\sim$8 kpc. Similarly, most of the OB stars (60 out of 69 stars) in the \cite{daflon2004} sample are also within $\sim$ 9 Kpc from the galactic center; the flatter gradients obtained in that study may be influenced by this selection.

\begin{figure}
\centering
\resizebox{0.9\hsize}{!}{
\includegraphics{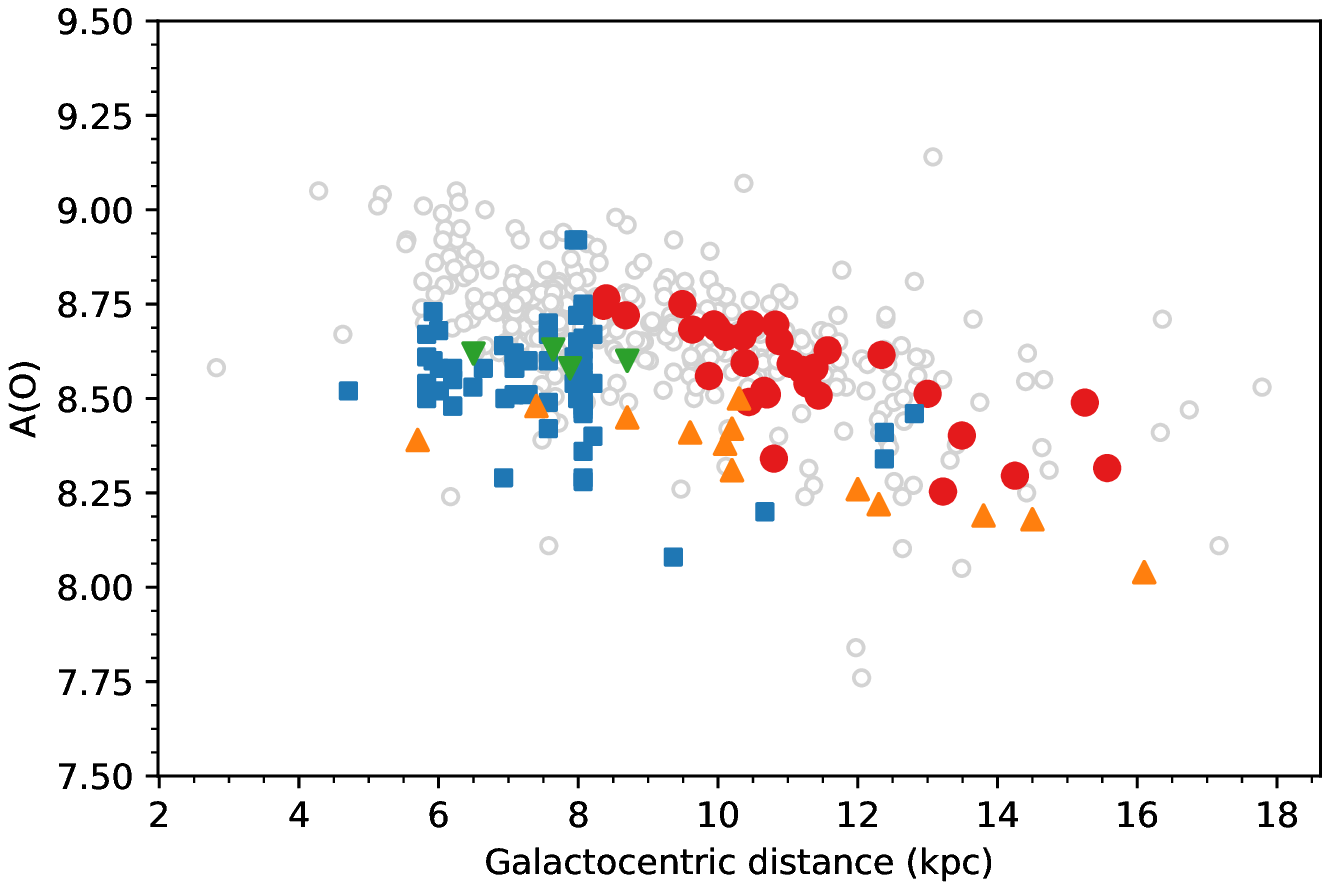}}
\caption{Radial gradient of oxygen abundance  traced by young objects of the Galactic disk: OB stars from  this work (red circles) and from \cite{daflon2004} (blue squares); cepheids from \cite{luck2018} (open circles); SFR regions from \cite{spina2017} (green triangles); and \ion{H}{ii} regions from \cite{esteban2017} (orange triangles).
	\label{fig:grad_all}
}
\end{figure}
Radial  gradients traced by young massive B stars (age $<$ 20 Myr)  are not affected by long-term processes such as  radial migration \citep{magrini2017}, for example. In Figure \ref{fig:grad_all} we compare the radial distributions of oxygen abundances obtained for different young objects of the Galaxy.  Abundances of main-sequence OB stars are from this work (red circles) and from \cite{daflon2004} (blue squares). Average oxygen abundances derived by \cite{luck2018} based on a multiphase analysis for a large sample of Cepheids are represented by gray open circles.  Nebular abundances of \ion{H}{ii} regions from \cite{esteban2017} are represented by orange triangles.  Oxygen abundances of young G- and K-type stars in star forming regions from \cite{spina2017} are shown as green triangles (the oxygen abundances were estimated from their [Fe/H] values). It is clear from Figure \ref{fig:grad_all} that the oxygen abundance pattern for different types of young objects in the Galactic  disk present an overall large spread. The results for Cepheids have significant scatter and overlap with the less scattered results for the OB stars derived in this study; the cepheids oxygen abundances, however, are offset from the results in \cite{daflon2004} for Rg $<$ $\sim$ 8 kpc. 
The \ion{H}{ii} region abundances have a smaller scatter but are systematically lower than the results for the OB stars and Cepheids.  A comparison between oxygen  abundances obtained for the Cocoon Nebula and its ionizing B stars by
\cite{grsde2014} shows a difference of $\sim$ 0.2 dex, the stellar abundances being higher than the nebular abundances  considering corrections due to depletion onto dust. Similar discrepancy has been  also discussed by \cite{sds2014} for the Orion nebula. Such  discrepancy in the star-gas abundances may be due to possible temperature fluctuations.

The stars in star forming regions follow the trend in \cite{daflon2004}. Although with systematic differences between the probes, when considering all results for the OB stars and \ion{H}{ii} regions, there is evidence for a possible flattening in the inner disk (Rg $<$ 8 kpc) and a steeper gradient between roughly Rg$\sim$ 8 -- 16 kpc, as seen for the open clusters. 

Chemical evolution models of the Galaxy by \cite{chiappini2001} and \cite{cescutti2007} considering two-infall episodes and inside-out formation of the Galactic disk predict negative radial abundance gradients. The halo surface mass density is the key parameter in their models that allows for the presence of gradients with different slopes: the adoption of a constant halo density within $\sim$ 8 kpc from the Galactic center, for example, produces flat gradients in the outer disk (Rg $>$ $\sim$16 Kpc). From our results, however, it is not possible to conclude if the gradients flatten in the outer disk, with a break in the metallicity slope at Rg $\sim$ 12 kpc.  In a simple exercise,  we split the sample stars in two groups, breaking at $R_G \sim 12.5$ kpc, and recomputed the gradients. The gradients in the two zones are consistent within the uncertainties to those obtained for the full sample. In addition, we note a larger abundance spread in the outermost zone, producing higher uncertainties in the recomputed slopes: the radial gradient of oxygen for $R_G<12.5$ kpc is $-0.06\pm0.02$ dex/kpc whereas in the outer zone we obtained a gradient of $-0.04\pm0.04$ dex/kpc.

The more sophisticated chemodynamical models  by \cite{minchev2013,minchev2014} predict steeper radial metallicity gradient of $-0.057$ dex/kpc  from a subsample of  young (age < 2 Gyr) stars located close to disk midplane ($|z| < $0.25 kpc), while the radial gradient tends to flatten down to $-0.020$ dex/kpc for subsamples with increasing  age and increasing distance to the plane.  Our gradients obtained from abundances of young, massive B stars are consistent with these predictions of  present-day radial metallicity gradient in the Galactic plane.

\section{Conclusions}
\label{sec:conclusion}

We present a detailed elemental abundance study of 28 main-sequence OB stars located towards the Galactic anti-center plus three stars in the solar neighborhood.  We developed a robust and self-consistent methodology to analyze the spectra of these objects using a full Non-LTE approach, meaning
that the stellar atmosphere models and synthetic spectra were both calculated in Non-LTE using the codes \texttt{TLUSTY} and \texttt{SYNSPEC}. The analytic tools used to compare the observed and synthetic spectra were built using IDL/GDL and Python technologies.  The outputs of our iterative scheme are effective temperature, surface gravity, line broadening parameters (projected rotational velocity, macroturbulent and microturbulent velocities) as well as the abundances of oxygen and silicon. This methodology has been tested for three OB stars with abundance results previously presented in the literature and our present results of stellar parameters and abundances are consistent with the published values.
The studied sample contains  main-sequence OB stars with temperatures in the range 20\,800 to 31\,300\,K. From the derived stellar parameters and evolutionary tracks,  the masses of our target stars  vary from 7\,M$_{\odot}$ to 20\,M$_{\odot}$, with most of the stars having masses between 9 and 15\,M$_{\odot}$.

With  the derived elemental abundances from Table~\ref{tab:sample_results} and heliocentric distances from Table~\ref{tab:distances}, we investigated the variation of oxygen and silicon abundances across the outer disk of the Milky Way. We found no evidence of an azimuthal gradient, although our sample is not suitable for such study, so further work is needed to confirm this. Regarding the radial gradient, we found  slopes of $-0.07 \pm0.01$ and $-0.09 \pm0.01$\,dex/kpc for oxygen and silicon, respectively, in the region of $8.4 <R_G <15.6$\,kpc. The negative slopes indicate the abundances systematically decrease from  values  roughly solar at the solar neighborhood to average values that are approximately 0.4$-$0.6\,dex lower at $R_G \sim 15$\,kpc. The gradients obtained present a general good agreement when compared to other results of the literature covering the outer disk. The slopes we obtained for our sample of Main Sequence OB  stars are consistent with those predicted by chemodynamical models for a subsample of  young stars located in the Galactic midplane.

\begin{acknowledgement}

We thank the referee for a careful reading of the manuscript.  
G.A.B acknowledges financial support from CAPES (Ph.D. fellowship) and CNPq (post-doctoral fellowship), and the Observatoire de la C\^ote d'Azur for the warming hospitality. T.B. was funded by the project grant ``The New Milky Way''
from Knut and Alice Wallenberg Foundation. M.S.O and T.B. were supported by the US National Science Foundation, grant
AST-0448900 to MSO.  K.C. acknowledges funding from NSF grant AST-907873.

This research has made use of NASA’s ADS Bibliographic Services, the SIMBAD database, operated by CDS, Strasbourg, Python, Astropy \cite{astropy2013}, IPython \cite{perezgranger2007}, Matplotlib \cite{hunter2007}, NumPy
\cite{vanderwalt2011}. This work has made use of data from the European Space Agency (ESA) mission {\it Gaia} (\url{https://www.cosmos.esa.int/gaia}), processed by the {\it Gaia} Data Processing and Analysis Consortium (DPAC,
\url{https://www.cosmos.esa.int/web/gaia/dpac/consortium}). Funding for the DPAC has been provided by national institutions, in particular the institutions participating in the {\it Gaia} Multilateral Agreement.

\end{acknowledgement}

\bibliographystyle{aa}
\bibliography{referenser}

\end{document}